\documentclass[
reprint,
 aps,
 superscriptaddress,
bibnotes,
pra
]{revtex4-2}
\usepackage{physics}
\usepackage{dsfont}
\usepackage{color,newfloat}
\usepackage{graphicx}
\usepackage{dcolumn}
\usepackage{bm}
\usepackage{amsmath,amssymb,amsthm, amsfonts}
\usepackage{mathtools}
\usepackage{multirow}
\usepackage{hhline}
\usepackage{comment}
\usepackage[dvipsnames]{xcolor}
\usepackage{hyperref}
\hypersetup{
    colorlinks=true,
    linkcolor=Maroon,
    citecolor=Maroon,
    urlcolor=Maroon
}
\usepackage{gensymb} 
\usepackage{placeins}

\usepackage[caption = false]{subfig}
\usepackage{lineno}
\usepackage{siunitx}
\usepackage[mathscr]{euscript}
 \let\mathscr\relax

\usepackage{makecell}

\begin{document}


\title{
Photon Sorting with a Quantum Emitter}

\author{Kasper H. Nielsen}
\thanks {These authors contributed equally to this work}
\affiliation{Center for Hybrid Quantum Networks (Hy-Q), Niels Bohr Institute, University of Copenhagen, Jagtvej 155A, Copenhagen 2200, Denmark}
\affiliation{NNF Quantum Computing Programme, Niels Bohr Institute, University of Copenhagen, Blegdamsvej 17, 2100 Copenhagen, Denmark.}

\author{Etienne Corminboeuf}
\thanks {These authors contributed equally to this work}
\affiliation{Center for Hybrid Quantum Networks (Hy-Q), Niels Bohr Institute, University of Copenhagen, Jagtvej 155A, Copenhagen 2200, Denmark}

\author{Benedikt Tissot}
\affiliation{Center for Hybrid Quantum Networks (Hy-Q), Niels Bohr Institute, University of Copenhagen, Jagtvej 155A, Copenhagen 2200, Denmark}

\author{Love A. Pettersson}
\affiliation{Center for Hybrid Quantum Networks (Hy-Q), Niels Bohr Institute, University of Copenhagen, Jagtvej 155A, Copenhagen 2200, Denmark}
\affiliation{NNF Quantum Computing Programme, Niels Bohr Institute, University of Copenhagen, Blegdamsvej 17, 2100 Copenhagen, Denmark.}

\author{Sven Scholz}
\affiliation{Lehrstuhl f{\"u}r Angewandte Festk{\"o}rperphysik, Ruhr-Universit{\"a}t Bochum, Universit{\"a}tsstrasse 150, D-44780 Bochum, Germany}

\author{Arne Ludwig}
\affiliation{Lehrstuhl f{\"u}r Angewandte Festk{\"o}rperphysik, Ruhr-Universit{\"a}t Bochum, Universit{\"a}tsstrasse 150, D-44780 Bochum, Germany}

\author{Leonardo Midolo}
\affiliation{Center for Hybrid Quantum Networks (Hy-Q), Niels Bohr Institute, University of Copenhagen, Jagtvej 155A, Copenhagen 2200, Denmark}

\author{Anders S. Sørensen}
\affiliation{Center for Hybrid Quantum Networks (Hy-Q), Niels Bohr Institute, University of Copenhagen, Jagtvej 155A, Copenhagen 2200, Denmark}

\author{Peter Lodahl}
\affiliation{Center for Hybrid Quantum Networks (Hy-Q), Niels Bohr Institute, University of Copenhagen, Jagtvej 155A, Copenhagen 2200, Denmark}

\author{Ying Wang}
\email {ying.wang@nbi.ku.dk}
\affiliation{Center for Hybrid Quantum Networks (Hy-Q), Niels Bohr Institute, University of Copenhagen, Jagtvej 155A, Copenhagen 2200, Denmark}

\author{Stefano Paesani}
\email{stefano.paesani@nbi.ku.dk}
\affiliation{Center for Hybrid Quantum Networks (Hy-Q), Niels Bohr Institute, University of Copenhagen, Jagtvej 155A, Copenhagen 2200, Denmark}
\affiliation{NNF Quantum Computing Programme, Niels Bohr Institute, University of Copenhagen, Blegdamsvej 17, 2100 Copenhagen, Denmark.}

\begin{abstract}
High-quality photonic Bell state measurements (BSMs) enable scalable universal quantum computing and long distance quantum communication. 
However, when implemented with linear optics, BSMs are fundamentally probabilistic, introducing substantial hardware overheads and limiting noise tolerance in photonic quantum computing architectures.
Nonlinear interactions at the single-photon level can overcome these limitations by enabling near-deterministic photon-photon gates.
Here, we demonstrate a passive photon-sorting circuit based on the induced nonlinearity arising from photon scattering in a solid-state quantum emitter. 
The scattering is implemented in a directional waveguide-emitter coupling interface and embedded on-chip into a linear optical circuit, through which we demonstrate sorting of one- and two-photon components with a success probability of $62\%$. 
We find that the current system can enable BSMs with a $57\%$ post-selected success probability without ancillary photons, exceeding the linear-optical limit of $50\%$, and can be readily improved to $>65\%$ with design optimisations. 

\end{abstract}

\date{April 24, 2026}

\maketitle


\section{Introduction}


Photons are an attractive means for developing quantum technologies, such as photonic quantum computers~\cite{knill2001ascheme,bartolucci_comparison_2025} and quantum networks~\cite{tiarks_optical_2016,duan_long-distance_2001}, due to their low decoherence and fast speed.
However, deterministic photon-photon operations are challenging to implement, as photons do not interact with one another~\cite{tiarks_photonphoton_2019,tiarks_photonphoton_2019,chang_quantum_2014,calsamiglia_maximum_2001}. 
This results in the lack of native deterministic two-qubit operations with linear optics~\cite{knill2001ascheme,sparrow2018simulating}. 
Photonic quantum computing architectures have been developed to support fault-tolerant systems even in the presence of probabilistic operations~\cite{knill2001ascheme,raussendorf2003measurement, bartolucci2023fusion}.
For example, fusion-based quantum computing (FBQC)~\cite{bartolucci2023fusion} relies on the preparation of many copies of small-scale entangled resource states of photons. These are then consumed by Bell state measurements (BSMs), often called fusions, that perform the computation. 
These BSMs are native in linear optics, but only succeed with 50\% probability \cite{calsamiglia_maximum_2001, weinfurter_experimental_1994, braunstein_measurement_1995}. 
The success chance can be boosted arbitrarily by adding ancillary photons, but the architectures become more susceptible to the loss of any of the photons involved~\cite{ewert2014efficient,grice2011arbitrarily,hauser2025boosted}. 
As a consequence, the loss tolerance of photonic architectures using boosted fusions is limited to $<0.8\%$~\cite{bartolucci2023fusion}. This tolerance can be increased to $<8\%$ with specific encodings ~\cite{chan2025tailoring, dessertaine_enhanced_2024}, which are challenging to achieve experimentally. Even larger loss thresholds can be achieved by increasing the size of the initial resource states, but at the cost of creating a large entangled photonic state~\cite{bartolucci_comparison_2025}, which is an experimentally demanding task.

The development of strong nonlinearities between the photons can be used to avoid these challenges~\cite{chang_quantum_2014}. 
In particular, a photon sorter, i.e. a device capable of separating the one- and two-photon components in a light pulse, has been proposed as a key nonlinear functionality that can strongly enhance the success probability of BSMs~\cite{witthaut2012photon}. Furthermore, a recent theoretical study demonstrated reduced overhead and increased loss tolerance in a measurement-based-quantum computing architecture by introducing scattering-based nonlinear operations~\cite{ostmann2025nonlinear}. 
Several theoretical proposals for creating a photon sorter that utilises the nonlinearity of light-emitter interaction in combination with linear optical circuitry have been put forward~\cite{witthaut2012photon,yang2022deterministic,ralph2015photon}. 
However, the required strong nonlinear interactions at the few-photon level are challenging to achieve in the optical domain~\cite{ewaniuk2023imperfect,steinbrecher2019quantum,javadi2015single, birnbaum_photon_2005, hacker_photonphoton_2016, tiarks_optical_2016, tiarks_photonphoton_2019}.
One way to mediate a strong photon-photon interaction is to scatter photon pulses from a solid-state quantum emitter, such as a semiconductor quantum dot (QD) that is incorporated in a nanophotonic cavity or waveguide~\cite{uppu2021quantum}. 
The scattering interaction between a photon pulse and a quantum emitter has been studied in detail both theoretically~\cite{fan2010input,witthaut2012photon,yang2022deterministic,kiilerich2019input} and experimentally~\cite{javadi2015single,jeannic2022dynamical,Tomm2023,nielsen_programmable_2025, desantis2017asolid} in recent years. 
In fact, while proxies for the functionality of a photon sorter have been experimentally investigated before, they were based on a bidirectionally coupled QD, a configuration incompatible with near-deterministic operation~\cite{bennett2016asemiconductor}. 
Additionally, previous experiments analysed modifications to the autocorrelation functions of the incident light field, whereas photon-number distributions demonstrating actual photon-sorting behaviour were not directly observed. 
Furthermore, implementing such nonlinearities in a strong and near-deterministic regime necessitates unidirectional (e.g. chiral) coupling between the emitter and waveguide~\cite{lodahl2017chiral}.

Here, we present an experimental implementation of a photon sorter utilising an effectively chiral light-matter interaction between solid-state quantum emitters embedded in a nanophotonic waveguide (Fig.~\ref{fig:1}d) combined with a linear optical circuit. 
We measure the photon-number distribution of the output field and observe the sorting of the single- and two-photon components into two distinct spatial output modes (Fig.~\ref{fig:1}a), providing an unambiguous demonstration of a photon sorter. 
Furthermore, we demonstrate that the fidelity of the nonlinear photon sorter embedded into a BSM exceeds the bounds achievable with linear optics, thereby opening a pathway to nonlinearity-boosted photonic fusion.
Finally, we analyse how such enhancements would affect key quantum photonic applications, such as loss-tolerant quantum computing architectures and quantum repeaters, and show that significant practical advances can be achieved with state-of-the-art QD emitter technology. 

\section{Photon sorter implementation and performance}

\begin{figure}
    \centering
    \includegraphics{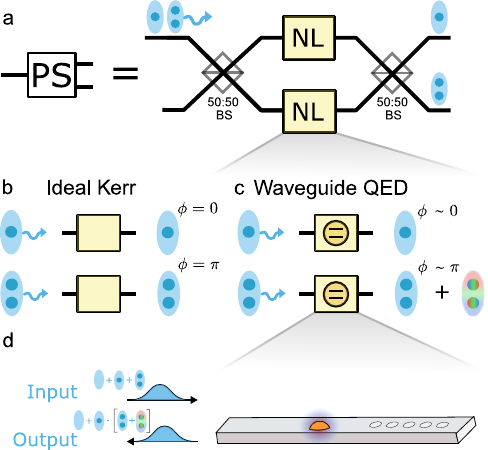}
    \caption{\textbf{Photon sorter building blocks.} 
    a. Sketch of the photon sorter implementation using a nonlinear photonic interaction between two balanced beam splitters (a nonlinear Mach-Zehnder interferometer).
    Ideally, individual photons will exit through the upper port, whereas the two-photon component will exit through the lower.
    b. Illustration of an ideal single-mode Kerr nonlinearity where a one-photon state acquires no phase shift, whereas a two-photon state receives a phase shift of $\pi$.
    c. Illustration of the nonlinear response of a two-level system. The photon states receive approximately the same phase as in the ideal Kerr case, but the interaction distorts the two-photon wavepacket and creates a superposition of the elastically scattered state with an inelastically scattered (entangled) component. 
    d. Schematically, the quantum emitter system consists of a QD coupled to a single-sided waveguide terminated by a mirror, effectively making the light-matter interaction chiral. 
}
    \label{fig:1}
\end{figure}

An ideal photon sorter is comprised of a balanced Mach-Zehnder Interferometer (MZI) consisting of two 50:50 beam splitters and a single-mode nonlinear element in both arms, as seen in Fig.~\ref{fig:1}a. 
This nonlinear element would ideally correspond to a Kerr nonlinearity, giving rise to the unitary evolution $U(k)=\exp(ik\pi\hat{n}(\hat{n}-1)/2)$, where $n$ represents the number of photons and $k$ defines the strength of the nonlinearity~\cite{steinbrecher2019quantum}. 
For $k=1$, this gives a $\pi$ phase shift to a two-photon state, with no phase shift incurred on a one-photon state, see Fig.~\ref{fig:1}b. 
Therefore, as sketched in Fig.~\ref{fig:1}a, such a configuration perfectly directs the one-photon component in the input state into the top output mode due to constructive interference ($\phi = 0$), and the two-photon component into the bottom mode due to destructive interference ($\phi = \pi$). 
Ref.~\cite{witthaut2012photon} introduced an approximate configuration that is experimentally practical, proposing the realisation of a Kerr-like nonlinearity using two identical two-level systems, see Fig.~\ref{fig:1}c.
 As a two-photon component is resonant with the two-level system, the scattering entails an elastic single-mode component alongside an inelastic multi-mode component~\cite{fan2010input}. This nonlinearity results in a phase shift whose deviation from the ideal Kerr case of $\pi$ depends on the photon-emitter interaction strength (see a detailed discussion in Appendix~\ref{appendix:theory}).

 We employ a single-sided nanobeam waveguide (Fig.~\ref{fig:1}d) to facilitate the QD-mediated photon-photon interaction. By terminating one end with a photonic crystal mirror, the system ensures that photons are scattered predominantly in a single direction. This system effectively mimics a true chiral waveguide~\cite{lodahl2017chiral,sollner2015deterministic,kirsanske2017indistinguishable}, but is significantly more robust experimentally (see Appendix~\ref{appendix:theory} for a comparison between single-sided and chiral waveguides). 
The QD coupling efficiency to the waveguide mode is defined as $\beta = \gamma / (\gamma + \gamma_{\mathrm{loss}})$, representing the fraction of QD emission funnelled into the guided mode. $\gamma_{\mathrm{loss}}$ denotes the decay rate into undesired radiation modes, which depends on the spatial position of the QD. 
To suppress charge noise experienced by the QD in the host material, a p-i-n diode is utilised to facilitate electrical gating. 
Crucially, the metal gate enables DC Stark tuning of the QD resonance frequency, serving as an additional control knob to modulate the QD-mediated photon-photon interactions.
Furthermore, an on-chip beam splitter is integrated to spatially separate the input and output fields, thereby substantially simplifying the experimental implementation and suppressing reflections from the coupler's surface (see Appendix~\ref{appendix:expedetails}). 
To realise optimal QD-mediated interactions, Gaussian pulses with a desired duration of approximately twice the QD lifetime are carved from a continuous-wave (CW) laser~\cite{witthaut2012photon,jeannic2022dynamical}.
This field is subsequently attenuated to a mean photon number of $\overline{n} \approx 0.1$ within the pulse, ensuring that the input is primarily a superposition of the zero-, one- and two-photon Fock states.

\begin{figure*}
    \centering
    \includegraphics[width=\textwidth]{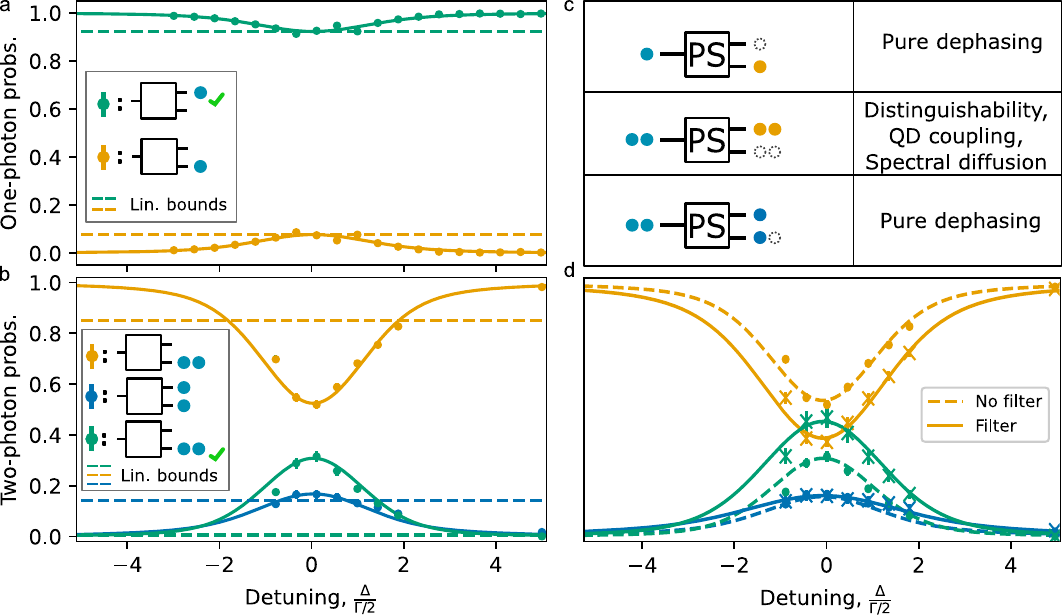}
    \caption{\textbf{Photon sorter performance characterisation.}
    a. Probabilities of detecting the one-photon state in the upper (green) or lower (orange) output port of the photon sorter as a function of the detuning $\Delta$ between the input field and the QD, normalised to the QD linewidth $\Gamma/2$.  
    The dashed lines correspond to a linear optical beam splitter with a reflectivity matching the one-photon statistics at $\Delta=0$.
    b. Probabilities of detecting two photons in the upper mode (orange), two photons in the lower mode (green) or one photon each in both modes (blue) as a function of the detuning $\Delta$. 
    An ideal photon sorter would transfer the two photons into the lower mode completely.
    The dashed lines correspond to the two-photon statistics from the beam splitter calibrated to the one-photon statistics at $\Delta=0$. The enhanced performance of the nonlinear photon sorter relative to a linear optical element is evident.
    c. Mechanisms leading to incorrect sorting of one- and two-photon states. 
    For single photons, errors arise solely from pure dephasing.
    For two photons exiting in the upper mode, an imperfect Kerr nonlinearity distorts the wavepacket, thereby reducing photon indistinguishability. Additionally, a non-unity QD-waveguide coupling $\beta$ and spectral diffusion further increase this probability. When two photons exit through different ports, the error originates only from pure dephasing.
    d. Implementation of temporal filtering of the two-photon events. Filtering increases the effective coupling $\beta$. As a result, the probability that two-photon states are sorted into the correct mode increases, whereas the probability of the two photons splitting up into two paths remains similar.
    In all plots, experimental data are represented by markers,  while theoretical fits are the associated solid lines. 
    }
    \label{fig:2}
\end{figure*}

Incorporating this nonlinear block into the photon sorter architecture (Fig.~\ref{fig:1}a), we experimentally adopt a time-bin protocol that allows sequential scattering from a single emitter rather than two distinct emitters.
This approach overcomes the fundamental challenge of growing strictly identical QDs, thereby eliminating operational errors induced by inherent emitter inhomogeneity. 
Moreover, the implementation of a double-pass time-bin interferometer provides passive self-phase stabilisation.
In this configuration, the first pass encodes the input pulses into time bins, while the second path re-interferes the post-interaction photons, executing the photon-sorting operation by mapping temporal modes onto spatial modes (see details in Appendix~\ref{appendix:expedetails}).
Finally, pseudo-photon-number-resolving detection is performed to evaluate the capability of the photon sorter.

Results of the one-photon statistics, which correspond to normalised intensity measurements, are shown in Fig.~\ref{fig:2}a as a function of the detuning $\Delta$.
As we approach resonance, the intensity of light in the second mode increases, a behaviour that cannot be reproduced assuming identical QD scattering in both time bins. This contribution arises from a fast noise term that occurs on a timescale shorter than the 5\,ns delay between the early and late time bins, making the two subsequent QD interactions slightly different. We believe that this is caused by phonon-induced pure dephasing, a mechanism that also contributes to partial distinguishability from a QD single-photon source \cite{lodahl2015interfacing}.
The two-photon statistics is shown in Fig.~\ref{fig:2}b. Far from resonance, as expected, the photons remain in the first mode since the non-linear interaction with the QD is negligible. As we approach resonance, it becomes increasingly likely that the photons are transferred into the second mode.  

For the one-photon statistics, at $\Delta = 0$ we find a lower bound of our photon sorter to correctly sort the one-photon state as $P_{10} = 92\% \pm 1\%$. We can select a linear circuit of a variable beam splitter yielding the same output distribution (see dashed lines in Fig.~\ref{fig:2}a). For two-photon statistics, at $\Delta = 0$, the photon sorter correctly sorts two-photon states maximally with a probability of $P_{02} \approx 32\pm2\%$. The identical linear circuit selected above enables a success probability of only $P_{02}^{lin} \approx 0.64\%$ (dashed lines Fig.~\ref{fig:2}b). Evidently, our photon sorter significantly outperforms its linear counterpart in this scenario. 

The overall success probability of the photon sorter is given by the average success probability over one- and two-photon states
\begin{equation}
    P=\frac{1}{2}(P_{10}+P_{02})=62\%\pm2\%,
\end{equation}
compared to the ideal linear optical bound of 50\%, which is achieved by either an identity or swap operation (i.e. not sorting the photons at all).

The mechanisms that induce incorrect two-photon state-sorting behaviours are distinct. The splitting of the two-photon state into one photon in each output, shown in Fig.~\ref{fig:2}b (blue) with a probability of $16 \%$, is not expected from a photon sorter with identical emitters in both modes. Phonon-induced pure dephasing can explain this effect. Regarding the two photons routed to the upper mode (Fig.~\ref{fig:2}b (orange)), this can be attributed to a non-unity QD-waveguide coupling efficiency ($\beta$), which lowers the nonlinear interaction strength. Spectral diffusion plays a similar role, reducing the coupling to the QD by inducing a finite (stochastic) detuning from resonance. Additionally, inelastic scattering distorts the photon wavepacket, rendering the inelastic two-photon state partially distinguishable from its elastic counterpart.
The mechanisms of noise terms on the one- and two-photon statistics are further summarised in Fig.~\ref{fig:2}c.
\begin{figure*}
    \centering
    \includegraphics[width=
\textwidth]{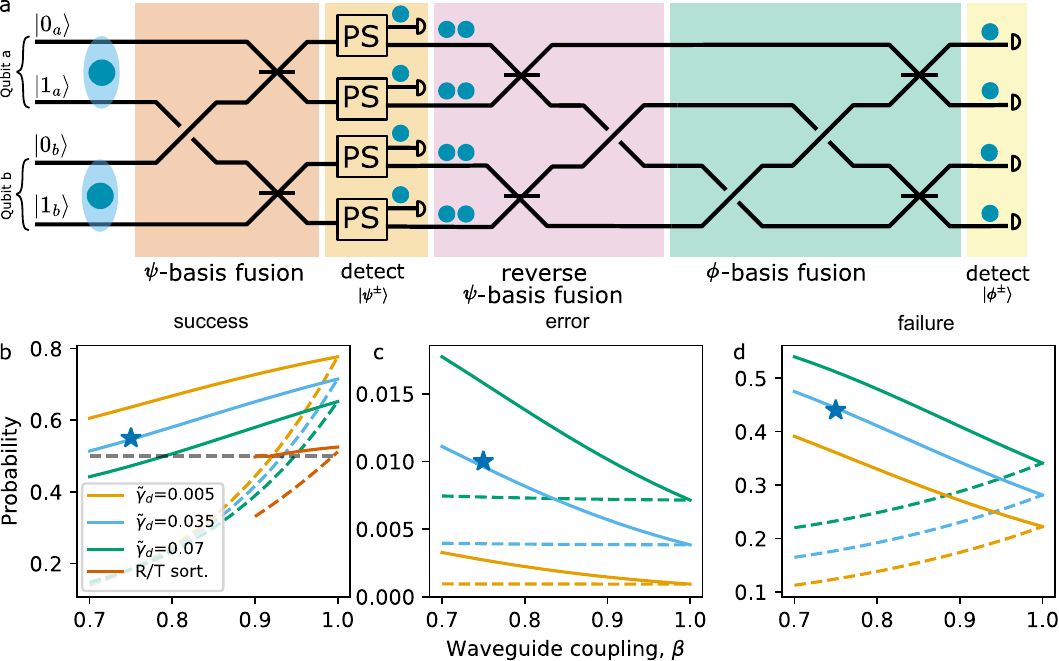}
    \caption{
    \textbf{Photon sorter for BSMs.}
     a. Illustration of a nonlinear BSMs using photon sorters. The first block (in orange) corresponds to a linear optical $\ket{\psi}$-fusion.
     Photons from the $\ket{\psi^\pm}$ states are incident individually on the photon sorter, so a direct measurement follows (yellow).  
     Photons from $\ket{\phi^\pm}$ bunch after the first fusion block and are split away by the photon sorters. Consecutively, a reversed-$\ket{\psi}$-fusion (pink) and a $\ket{\phi}$-basis fusion (green) are performed.
     We evaluate the performance of photon-sorter-boosted BSMs by analysing their b. success, c. error, and d. failure probabilities. Here, failures denote detected measurement patterns unassociated with any valid state, while errors correspond to misattributed states. 
     We model the associated probabilities, taking into account limitations such as the waveguide coupling of the emitter $\beta$, pure dephasing $\tilde \gamma_d$ and spectral diffusion $\tilde{\sigma}_\text{sd}$. We show the performance as a function of $\beta$ and $\tilde \gamma_d$, fixing the spectral diffusion at our experimentally measured value of $\tilde{\sigma}_\text{sd}=0.67$. 
     The dashed lines indicate the probabilities with waveguide losses included, whereas the solid lines are renormalised to account for losses, making them comparable to the experimental scenario. 
     The blue line corresponds to the pure dephasing rate measured in our system, and the star marks the actual probability achieved using our photon sorter.
     Finally, for comparison, the orange curves represent the alternative photon sorter architecture from Ref.~\cite{bennett2016asemiconductor}, which uses a two-level system in a two-sided waveguide to separate one- and two-photon states into reflected and transmitted fields. The results are simulated under ideal conditions without pure dephasing and spectral diffusion.}
    \label{fig:3}
\end{figure*}
The scattering dynamics with spectral diffusion and pure dephasing is modelled using input-output theory, and the states are propagated through the interferometer to find the theoretical output photon statistics. These are the solid lines in Fig.~\ref{fig:2}a, c. Details of the theoretical model are described in Appendix~\ref{appendix:theory}. 
The one- and two-photon statistics are fitted jointly to this model. From the photon statistics, the fit gives the emitter's waveguide coupling $\beta=0.75\pm0.01$ and the pure dephasing rate $\tilde{\gamma}_d=\frac{\gamma_d}{\Gamma/2}=0.035\pm0.001$.
Given that spectral diffusion occurs on a timescale much slower than the interferometer's 5\,ns time delay, we can model this effect by sampling detunings from a Gaussian distribution and find an experimental value from the fit of $\tilde{\sigma}_\text{sd}=\frac{\sigma_\text{sd}}{\Gamma/2}=0.67\pm0.07$.
%
It is noted that $\beta$ could be readily improved to near-unity by using a photonic crystal waveguide rather than a nanobeam waveguide \cite{arcari2014near}, which, assuming $\tilde \gamma_d, \tilde \sigma_{sd}$ from above, would increase the two-photon sorting probability  $P_{02}$ to 52.5\%. 

We apply a temporal filter to the output states to increase the fraction of inelastically scattered photons. Specifically, because of different temporal wavefunctions of the elastically and inelastically scattered two-photon states, we are able to improve the sorting probability by selecting photons more likely to belong to the inelastically scattered part. We temporally filter by selecting only two-photon detection counts within a $2.5\text{ns} \times 2.5 \text{ns}$ window.
This post-selection effectively increases $\beta$, which enhances the nonlinear interactions. On the other hand, the filtering reduces overall efficiency and induces losses, potentially making it undesirable for any application of a photon sorter that requires high efficiency. Details can be found in Appendix~\ref{appendix:filtering}.
The two-photon sorting probability is enhanced to $47 \%$  from $32 \%$ shown in Fig.~\ref{fig:2}d, where $>72\%$ \cite{witthaut2012photon} is reachable under ideal conditions of unity $\beta$, zero spectral diffusion, and zero pure dephasing.
\section{Photon sorter for Bell state measurements}

One of the main applications for photon sorting is to improve the success probability of BSMs.  
A linear optical BSM can only distinguish two of the four Bell states $\ket{{\psi^\pm}}=\frac{1}{\sqrt{2}}(\ket{01}\pm\ket{10}), \ket{{\phi^\pm}}=\frac{1}{\sqrt{2}}(\ket{00}\pm\ket{11})$ and thus has a 50\% success chance without boosting \cite{calsamiglia_maximum_2001}.
The layout of a proposed full Bell state analyser with photon sorters is shown in Fig.~\ref{fig:3}a. It first consists of a linear optical BSM, followed by a layer of photon sorters on each of the four modes. 
Since $\ket{\psi^\pm}$ states only have at most one photon per path, they can be fed directly to the detectors after passing through the photon sorter and are thus detected as usual in the linear BSM. 
Photons in $\ket{\phi^\pm}$ Bell states bunch before the photon-sorter layer, and give rise to non-unique detection patterns if measured directly. To address this, the photon sorter separates them from the $\ket{\psi^\pm}$ states. Subsequently, a layer of linear optics that reverses the $\ket{\psi^\pm}$-basis fusion is applied, and a $\ket{\phi^\pm}$-basis fusion is performed to enable unambiguous detection of all four Bell states.

Assuming perfect linear optics within the BSM, we evaluate the architecture's performance. While an ideal photon sorter \cite{yang2022deterministic} enables unit fidelity of the BSM, our architecture is limited to a success probability of $86\%$ in the case of a perfect, noiseless case. Sorter imperfections, such as imperfect waveguide coupling, pure dephasing and spectral diffusion of the QD reduce the success probability of the subsequent BSM.
To capture the analyser's behaviour under these limitations, we can classify the outcomes into three specific events, see Appendix~\ref{appendix:fusion} for details. 
i) Success (Fig.~\ref{fig:3}b): The Bell state is correctly identified. This occurs if all photons in the Bell state are correctly sorted and detected. Note that for $\ket{\phi^\pm}$ states, sorter-induced photon distinguishability due to pure dephasing of the emitter reduces the subsequent BSM success probability and overall fidelity.
ii) Error (Fig.~\ref{fig:3}c): The Bell state is misclassified as another Bell state. This occurs due to pure dephasing of the quantum emitter, making the photons exiting the photon sorter distinguishable. These distinguishable photons in a linear BSM are equally likely to be detected in any mode, sometimes resulting in a click pattern associated with another Bell state. 
iii) Failure (Fig.~\ref{fig:3}d): The BSM yields an identifiably wrong measurement pattern. This arises either from initial sorting errors or from the misidentification of correctly sorted states due to pure dephasing.
%

Note that to facilitate comparison between experimental measurements and the simulation, probabilities are based on the normalised performance of the photon sorter detection, meaning that any loss associated with the photon sorter is removed by post-selection (solid lines in the plots). Probabilities taking waveguide losses into account are shown as well (dashed lines). 
The success, error, and failure probabilities depend on $\beta$, as well as $\tilde{\gamma}_d$ and $\tilde{\sigma}_{sd}$. The dependence of these probabilities on $\beta$ and $\tilde{\gamma}_d$ is shown in Fig. \ref{fig:3}b,c,d, respectively, using the experimental spectral diffusion rate measured above.
The solid blue curve corresponds to the value of $\tilde \gamma_d$ measured for the sorter in this experiment. For $\beta=0.75$ obtained for the photon sorter above, the photon-sorter-boosted nonlinear BSM yields a normalised success rate of $57\%\pm2\%$, see Fig.~\ref{fig:3}b.
This value, marked by a star in the plots, is in line with the simulated value. Crucially, it surpasses the 50\% success limit inherent to linear optics, but remains below the ideal noiseless limit of 86\% \cite{witthaut2012photon} for this architecture.
This shortfall is primarily due to the low $\beta$ factor, which reduces the nonlinear photon-photon interaction (Fig.~\ref{fig:3}d). 
Evidently, increasing $\beta$ dramatically improves the BSM performance. Achieving a $\beta=0.98$, a value already reported in photonic crystal waveguides~\cite{arcari2014near}, would readily enhance the normalised BSM success probability beyond 64\% assuming similar pure dephasing and spectral diffusion rates as in the experiment. 
In the case where waveguide coupling losses are explicitly taken into account, Fig~\ref{fig:3}b shows that $\beta>0.9$ is required to beat the 50\% success probability of linear optics. 
Additionally, suppression of $\tilde \gamma_d$ increases the BSMs performance, primarily due to a significant reduction in the failure probability (yellow curve in Fig.~\ref{fig:3}d).
Lastly, the fidelity of the boosted BSM is compared with that of an alternative photon sorter implementation utilising an individual QD coupled to a conventional bidirectional waveguide \cite{bennett2016asemiconductor}. The simulated performance of this alternative is shown by the orange curve in Fig.~\ref{fig:3}b. The fidelity is evaluated as a function of $\beta$ assuming a noiseless QD in the absence of pure dephasing and spectral diffusion. As $\beta$ approaches unity, the success probability of this alternative approach extends marginally beyond the linear bound. 

\section{Advances for quantum applications}
\begin{figure}
    \centering
    \includegraphics[width=\linewidth]{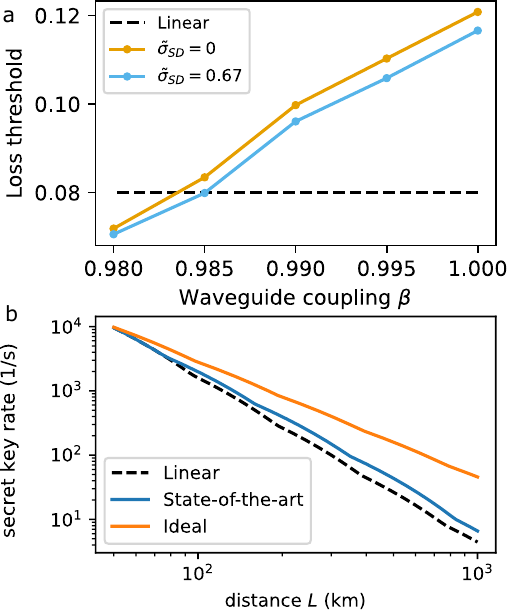}
    \caption{
    \textbf{Advances towards applications.}
    a. Loss threshold of the FBQC architecture introduced in Ref.~\cite{chan2025tailoring}, when using photon sorting to boost fusion performance, as a function of $\beta$. The two curves assume no pure dephasing, with and without spectral diffusion. The black dashed line denotes the threshold using linear optical BSMs.
    b. Performance of a repeater network for QKD using the DLCZ protocol~\cite{duan_long-distance_2001}, comparing linear optical BSMs (black dashed line),  and photon-sorter-boosted nonlinear BSMs.
    The latter includes a state-of-the-art scenario achieved solely by improving the $\beta=0.98$~\cite{arcari2014near} (blue), and an ideal noiseless photon sorter~\cite{witthaut2012photon} (orange).
    During initial entanglement generation, we assume that each link performs 100 attempts in parallel between individual memory qubits, conditionally emitting a photon per qubit.
    }
    \label{fig:4}
\end{figure}

BSMs are essential building blocks for photonic technology. Consequently, our proposed architecture, shown in Fig.~\ref{fig:3}, directly enables two main applications, as detailed in Fig.~\ref{fig:4}. 

In Fig.~\ref{fig:4}a, we report the loss threshold of the FBQC architecture introduced in Ref. \cite{chan2025tailoring} with fusion measurements replaced by a nonlinear BSM. The black dashed line corresponds to the threshold for a linear optical BSM with no photon sorting or boosting; the orange line corresponds to introducing a noiseless photon sorter; and the blue line corresponds to a photon sorter with spectral diffusion comparable to that of the emitter used in the experiment. The loss tolerance increases from 8\% to 12\% for the ideal case, but a waveguide coupling $\beta>0.985$ is required to gain any advantage of using a photon sorter. In both cases, the pure dephasing rate is assumed to vanish, $\gamma_d = 0$.

In Fig.~\ref{fig:4}b, we present the extrapolated secret-key rate for quantum key distribution (QKD) in a repeater network based on an ensemble-based repeater protocol \cite{duan_long-distance_2001,sangouard_quantum_2011,scarani_security_2009,bennett_mixed-state_1996}. 
Here, entanglement is generated between repeater stations, followed by the use of BSMs to extend the link range.
The resulting long-distance entanglement link can then be used to generate a secret key between the corresponding end nodes.
Thus, increasing the BSM success probability benefits the secret key rate.
We compare the rate for an all-linear BSM, a nonlinear BSM with a state-of-the-art QD (waveguide coupling from \cite{arcari2014near} and $\tilde \gamma_d, \tilde \sigma_{sd}$ from above), and an ideal, noiseless implementation \cite{witthaut2012photon} of a nonlinear BSM. We see that the state-of-the-art and the ideal nonlinear BSM significantly outperform the linear implementation. See Appendix~\ref{appendix:qkd} for more details.
For simplicity, we neglect the device losses associated with coupling the photonic memory to the photonic BSM, which is necessary in all the approaches.
These losses immediately reduce the swap success probability and, in turn, the total rate, underlining the importance of engineering efficient memory-BSM couplings.
Note that we explicitly take losses associated with imperfect waveguide coupling of the emitter into account in the estimations for both applications.
Based on these estimates, FBQC could benefit from emitter-based boosting of the fusion operation but would require state-of-the-art performance (coupling efficiency, ultralow noise and dephasing) to improve existing architectures and thresholds. It remains an interesting topic for future research to further engineer the FBQC architecture to fully exploit the capabilities of nonlinear BSMs. In repeater networks, state-of-the-art emitters can increase the secret key rate; thus, this application can already benefit from existing quantum emitters.
\section{Conclusion and outlook}
%
We demonstrate a passive photon sorter architecture that leverages the deterministic optical nonlinearity induced by scattering off a quantum emitter. The quantum emitter functions as an effective nonlinear phase shifter within a Mach-Zehnder interferometer, which enables the device to route one- and two-photon states from incident pulses to distinct spatial output modes. We measure a photon-sorting success probability of 62\%. One of the main applications of a photon-sorting device is to enhance BSMs, which are inherently limited to a $50\%$ success rate without boosting. 
Despite emitter couplings far below state-of-the-art levels, we find extracted BSM fidelities that exceed the $50\%$ threshold of linear optics upon post-selection, with an estimated BSM success rate of 57\%. The performance is robust to realistic dephasing noise, with state-of-the-art waveguide couplings enabling estimated fusion success probabilities exceeding $65\%$, which is comparable to two-photon boosting schemes \cite{ewert2014efficient,hauser2025boosted}.
The broader impact of this architecture is reflected in its potential to enhance FBQC and repeater networks. State-of-the-art quantum emitters can directly improve the performance of repeater networks. Substantial improvements to the loss threshold of FBQC can be achieved with further photonic engineering of the solid state environment of the quantum emitter.

Compared to previous photon sorter architectures in Ref.~\cite{bennett2016asemiconductor}, which measure the modified autocorrelation function induced by nonlinear interaction with a quantum emitter, we measure up to two-photon statistics in both output modes of the device. This is necessary to demonstrate performance for applications such as BSMs and to verify the photon-sorting device's output statistics. Additionally, we show that a photon sorter based on reflected and transmitted fields can only surpass the 50\% bound of linear optics by a few percent in an ideal noiseless implementation. Here, we utilise a single-sided device to create an effective chiral coupling between the emitter and waveguide mode, which is experimentally simpler than using a real chiral system \cite{sollner2015deterministic}.
An interesting path for future research was introduced in \cite{yang2022deterministic}, proposing an alternative approach to photon sorting which can achieve a success probability of $>99\%$. While still relying on scattering photon pulses to induce a nonlinear effect, it requires two scatterings and temporal-mode-shape filtering, significantly increasing the experimental complexity. 
%

The approach employed in this article lays the groundwork for quantum technologies that require near-deterministic entangling gates and enables the passive boosting of fusion success probability. With improved device performance, the photon-sorting setup demonstrated in this work thus opens a path to near-deterministic BSMs with promising applications in quantum information processing.

\ \\
\noindent\textbf{Acknowledgements}
We are grateful to Klaus Mølmer and Edward C.R. Deacon for fruitful discussions and Andreas Wieck for providing the equipment to grow the nanostructures.
We acknowledge funding from the Danish National Research Foundation (Center of Excellence “Hy-Q,” grant number DNRF139), the Novo Nordisk Foundation (Challenge project "Solid-Q"), the European Union’s Horizon 2020 research and innovation program under Grant Agreement No. 820445 (project name Quantum Internet Alliance). S.P. acknowledges funding from VILLUM FONDEN (MapQP, No. VIL60743), the European Research Council (ERC StG ASPEQT, No. 101221875), Danmarks Innovationsfond
research grant No. 4356-00009B (HyperTenQ) and from the NNF Quantum Computing Programme.
\ \\
\noindent \textbf{Author contributions}
K.H.N., E.C. and Y.W. carried out the experiment. Y.W. and L.M. designed and fabricated the nanostructures. S.S. and A.L. grew the sample. K.H.N. and E.C. analysed the data. B.T. and L.P modelled applications. P.L., S.P., Y.W. and A.S.S. supervised all aspects of the project. K.H.N. and E.C. wrote the manuscript with input from all authors.
\ \\
\noindent\textbf{Data Availability}
The data supporting the plots in this article and other findings of this study are available from the corresponding authors upon reasonable request.
\ \\
\noindent\textbf{Code Availability}
The code used for data analysis and numerical simulations is available from the corresponding authors upon reasonable request.
\ \\
\noindent\textbf{Conflicts of interest}
P.L. is a founder of the company Sparrow Quantum, which commercialises single-photon sources. The remaining authors declare no other conflicts of interest. 


\bibliography{bib}

@article{ramos2018correlated,
doi = {10.1088/1367-2630/aae73b},
url = {https://dx.doi.org/10.1088/1367-2630/aae73b},
year = {2018},
month = {oct},
publisher = {IOP Publishing},
volume = {20},
number = {10},
pages = {105007},
author = {Ramos, Tomás and García-Ripoll, Juan José},
title = {Correlated dephasing noise in single-photon scattering},
journal = {New Journal of Physics},}

@article{nielsen_programmable_2025,
	title = {Programmable nonlinear quantum photonic circuits},
	volume = {16},
	copyright = {2025 The Author(s)},
	issn = {2041-1723},
	url = {https://www.nature.com/articles/s41467-025-66205-w},
	doi = {10.1038/s41467-025-66205-w},
	number = {1},
	urldate = {2026-03-06},
	journal = {Nature Communications},
	publisher = {Nature Publishing Group},
	author = {Nielsen, Kasper H. and Wang, Ying and Deacon, Edward C. R. and Sund, Patrik I. and Liu, Zhe and Scholz, Sven and Wieck, Andreas D. and Ludwig, Arne and Midolo, Leonardo and Sørensen, Anders S. and Paesani, Stefano and Lodahl, Peter},
	month = dec,
	year = {2025},
	pages = {11397},
}

@article{witthaut2012photon,
doi = {10.1209/0295-5075/97/50007},
url = {https://doi.org/10.1209/0295-5075/97/50007},
year = {2012},
month = {mar},
publisher = {},
volume = {97},
number = {5},
pages = {50007},
author = {Witthaut, D. and Lukin, M. D. and Sørensen, A. S.},
title = {Photon sorters and QND detectors using single photon emitters},
journal = {Europhysics Letters}
}

@article{fan2010input,
  title = {Input-output formalism for few-photon transport in one-dimensional nanophotonic waveguides coupled to a qubit},
  author = {Fan, Shanhui and Kocaba\ifmmode \mbox{\c{s}}\else \c{s}\fi{}, S\"ukr\"u and Shen, Jung-Tsung},
  journal = {Phys. Rev. A},
  volume = {82},
  issue = {6},
  pages = {063821},
  numpages = {9},
  year = {2010},
  month = {Dec},
  publisher = {American Physical Society},
  doi = {10.1103/PhysRevA.82.063821},
  url = {https://link.aps.org/doi/10.1103/PhysRevA.82.063821}
}

@article{jeannic2022dynamical,
	author = {Le Jeannic, Hanna and Tiranov, Alexey and Carolan, Jacques and Ramos, Tom{\'a}s and Wang, Ying and Appel, Martin Hayhurst and Scholz, Sven and Wieck, Andreas D. and Ludwig, Arne and Rotenberg, Nir and Midolo, Leonardo and Garc{\'\i}a-Ripoll, Juan Jos{\'e} and S{\o}rensen, Anders S. and Lodahl, Peter},
	journal = {Nature Physics},
	number = {10},
	pages = {1191--1195},
	title = {Dynamical photon--photon interaction mediated by a quantum emitter},
	volume = {18},
	year = {2022}}

@article{yang2022deterministic,
  title = {Deterministic Photon Sorting in Waveguide QED Systems},
  author = {Yang, Fan and Lund, Mads M. and Pohl, Thomas and Lodahl, Peter and M\o{}lmer, Klaus},
  journal = {Phys. Rev. Lett.},
  volume = {128},
  issue = {21},
  pages = {213603},
  numpages = {6},
  year = {2022},
  month = {May},
  publisher = {American Physical Society},
  doi = {10.1103/PhysRevLett.128.213603},
  url = {https://link.aps.org/doi/10.1103/PhysRevLett.128.213603}
}

@article{hauser2025boosted,
author={Hauser, Nico
and Bayerbach, Matthias J.
and D'Aurelio, Simone E.
and Weber, Raphael
and Santandrea, Matteo
and Kumar, Shreya P.
and Dhand, Ish
and Barz, Stefanie},
title={Boosted Bell-state measurements for photonic quantum computation},
journal={npj Quantum Information},
year={2025},
month={Mar},
day={08},
volume={11},
number={1},
pages={41},
issn={2056-6387},
doi={10.1038/s41534-025-00986-2},
url={https://doi.org/10.1038/s41534-025-00986-2}
}

@article{ewert2014efficient,
  title = {$3/4$-Efficient Bell Measurement with Passive Linear Optics and Unentangled Ancillae},
  author = {Ewert, Fabian and van Loock, Peter},
  journal = {Phys. Rev. Lett.},
  volume = {113},
  issue = {14},
  pages = {140403},
  numpages = {5},
  year = {2014},
  month = {Sep},
  publisher = {American Physical Society},
  doi = {10.1103/PhysRevLett.113.140403},
  url = {https://link.aps.org/doi/10.1103/PhysRevLett.113.140403}
}

@article{grice2011arbitrarily,
  title = {Arbitrarily complete Bell-state measurement using only linear optical elements},
  author = {Grice, W. P.},
  journal = {Phys. Rev. A},
  volume = {84},
  issue = {4},
  pages = {042331},
  numpages = {6},
  year = {2011},
  month = {Oct},
  publisher = {American Physical Society},
  doi = {10.1103/PhysRevA.84.042331},
  url = {https://link.aps.org/doi/10.1103/PhysRevA.84.042331}
}

@article{ewaniuk2023imperfect,
author = {Ewaniuk, Jacob and Carolan, Jacques and Shastri, Bhavin J. and Rotenberg, Nir},
title = {Imperfect Quantum Photonic Neural Networks},
journal = {Advanced Quantum Technologies},
volume = {6},
number = {3},
pages = {2200125},
doi = {https://doi.org/10.1002/qute.202200125},
year = {2023}
}

@article{uppu2021quantum,
author={Uppu, Ravitej
and Midolo, Leonardo
and Zhou, Xiaoyan
and Carolan, Jacques
and Lodahl, Peter},
title={Quantum-dot-based deterministic photon--emitter interfaces for scalable photonic quantum technology},
journal={Nature Nanotechnology},
year={2021},
month={Dec},
day={01},
volume={16},
number={12},
pages={1308-1317},
issn={1748-3395},
doi={10.1038/s41565-021-00965-6},
url={https://doi.org/10.1038/s41565-021-00965-6}
}

@article{kiilerich2019input,
  title = {Input-Output Theory with Quantum Pulses},
  author = {Kiilerich, Alexander Holm and M\o{}lmer, Klaus},
  journal = {Phys. Rev. Lett.},
  volume = {123},
  issue = {12},
  pages = {123604},
  numpages = {6},
  year = {2019},
  month = {Sep},
  publisher = {American Physical Society},
  doi = {10.1103/PhysRevLett.123.123604},
  url = {https://link.aps.org/doi/10.1103/PhysRevLett.123.123604}
}

@article{steinbrecher2019quantum,
author={Steinbrecher, Gregory R.
and Olson, Jonathan P.
and Englund, Dirk
and Carolan, Jacques},
title={Quantum optical neural networks},
journal={npj Quantum Information},
year={2019},
month={Jul},
day={17},
volume={5},
number={1},
pages={60},
issn={2056-6387},
doi={10.1038/s41534-019-0174-7},
url={https://doi.org/10.1038/s41534-019-0174-7}
}

@article{javadi2015single,
author={Javadi, A.
and S{\"o}llner, I.
and Arcari, M.
and Hansen, S. Lindskov
and Midolo, L.
and Mahmoodian, S.
and Kir{\v{s}}ansk{\.{e}}, G.
and Pregnolato, T.
and Lee, E. H.
and Song, J. D.
and Stobbe, S.
and Lodahl, P.},
title={Single-photon non-linear optics with a quantum dot in a waveguide},
journal={Nature Communications},
year={2015},
month={Oct},
day={23},
volume={6},
number={1},
pages={8655},
issn={2041-1723},
doi={10.1038/ncomms9655},
url={https://doi.org/10.1038/ncomms9655}
}

@article{ralph2015photon,
  title = {Photon Sorting, Efficient Bell Measurements, and a Deterministic Controlled-$Z$ Gate Using a Passive Two-Level Nonlinearity},
  author = {Ralph, T. C. and S\"ollner, I. and Mahmoodian, S. and White, A. G. and Lodahl, P.},
  journal = {Phys. Rev. Lett.},
  volume = {114},
  issue = {17},
  pages = {173603},
  numpages = {5},
  year = {2015},
  month = {Apr},
  publisher = {American Physical Society},
  doi = {10.1103/PhysRevLett.114.173603},
  url = {https://link.aps.org/doi/10.1103/PhysRevLett.114.173603}
}

@Article{bennett2016asemiconductor,
author={Bennett, A. J.
and Lee, J. P.
and Ellis, D. J. P.
and Farrer, I.
and Ritchie, D. A.
and Shields, A. J.},
title={A semiconductor photon-sorter},
journal={Nature Nanotechnology},
year={2016},
month={Oct},
day={01},
volume={11},
number={10},
pages={857-860},
issn={1748-3395},
doi={10.1038/nnano.2016.113},
url={https://doi.org/10.1038/nnano.2016.113}
}

@article{desantis2017asolid,
author={De Santis, Lorenzo
and Ant{\'o}n, Carlos
and Reznychenko, Bogdan
and Somaschi, Niccolo
and Coppola, Guillaume
and Senellart, Jean
and G{\'o}mez, Carmen
and Lema{\^i}tre, Aristide
and Sagnes, Isabelle
and White, Andrew G.
and Lanco, Lo{\"i}c
and Auff{\`e}ves, Alexia
and Senellart, Pascale},
title={A solid-state single-photon filter},
journal={Nature Nanotechnology},
year={2017},
month={Jul},
day={01},
volume={12},
number={7},
pages={663-667},
issn={1748-3395},
doi={10.1038/nnano.2017.85},
url={https://doi.org/10.1038/nnano.2017.85}
}

@article{bartolucci2023fusion,
author={Bartolucci, Sara
and Birchall, Patrick
and Bomb{\'i}n, Hector
and Cable, Hugo
and Dawson, Chris
and Gimeno-Segovia, Mercedes
and Johnston, Eric
and Kieling, Konrad
and Nickerson, Naomi
and Pant, Mihir
and Pastawski, Fernando
and Rudolph, Terry
and Sparrow, Chris},
title={Fusion-based quantum computation},
journal={Nature Communications},
year={2023},
month={Feb},
day={17},
volume={14},
number={1},
pages={912},
issn={2041-1723},
doi={10.1038/s41467-023-36493-1},
url={https://doi.org/10.1038/s41467-023-36493-1}
}

@article{knill2001ascheme,
	author = {Knill, E. and Laflamme, R. and Milburn, G. J.},
	journal = {Nature},
	number = {6816},
	pages = {46--52},
	title = {A scheme for efficient quantum computation with linear optics},
	volume = {409},
	year = {2001}}

@article{raussendorf2003measurement,
  title = {Measurement-based quantum computation on cluster states},
  author = {Raussendorf, Robert and Browne, Daniel E. and Briegel, Hans J.},
  journal = {Phys. Rev. A},
  volume = {68},
  issue = {2},
  pages = {022312},
  numpages = {32},
  year = {2003},
  month = {Aug},
  publisher = {American Physical Society},
  doi = {10.1103/PhysRevA.68.022312},
  url = {https://link.aps.org/doi/10.1103/PhysRevA.68.022312}
}

@article{lodahl2017chiral,
	author = {Lodahl, Peter and Mahmoodian, Sahand and Stobbe, S{\o}ren and Rauschenbeutel, Arno and Schneeweiss, Philipp and Volz, J{\"u}rgen and Pichler, Hannes and Zoller, Peter},
	journal = {Nature},
	number = {7638},
	pages = {473--480},
	title = {Chiral quantum optics},
	volume = {541},
	year = {2017}}

@article{sollner2015deterministic,
	author = {S{\"o}llner, Immo and Mahmoodian, Sahand and Hansen, Sofie Lindskov and Midolo, Leonardo and Javadi, Alisa and Kir{\v s}ansk{\.e}, Gabija and Pregnolato, Tommaso and El-Ella, Haitham and Lee, Eun Hye and Song, Jin Dong and Stobbe, S{\o}ren and Lodahl, Peter},
	journal = {Nature Nanotechnology},
	number = {9},
	pages = {775--778},
	title = {Deterministic photon--emitter coupling in chiral photonic circuits},
	volume = {10},
	year = {2015}}

@article{kirsanske2017indistinguishable,
  title = {Indistinguishable and efficient single photons from a quantum dot in a planar nanobeam waveguide},
  author = {Kir\ifmmode \check{s}\else \v{s}\fi{}ansk\ifmmode \dot{e}\else \.{e}\fi{}, Gabija and Thyrrestrup, Henri and Daveau, Rapha\"el S. and Dree\ss{}en, Chris L. and Pregnolato, Tommaso and Midolo, Leonardo and Tighineanu, Petru and Javadi, Alisa and Stobbe, S\o{}ren and Schott, R\"udiger and Ludwig, Arne and Wieck, Andreas D. and Park, Suk In and Song, Jin D. and Kuhlmann, Andreas V. and S\"ollner, Immo and L\"obl, Matthias C. and Warburton, Richard J. and Lodahl, Peter},
  journal = {Phys. Rev. B},
  volume = {96},
  issue = {16},
  pages = {165306},
  numpages = {11},
  year = {2017},
  month = {Oct},
  publisher = {American Physical Society},
  doi = {10.1103/PhysRevB.96.165306},
  url = {https://link.aps.org/doi/10.1103/PhysRevB.96.165306}
}

@article{arcari2014near,
  title = {Near-Unity Coupling Efficiency of a Quantum Emitter to a Photonic Crystal Waveguide},
  author = {Arcari, M. and S\"ollner, I. and Javadi, A. and Lindskov Hansen, S. and Mahmoodian, S. and Liu, J. and Thyrrestrup, H. and Lee, E. H. and Song, J. D. and Stobbe, S. and Lodahl, P.},
  journal = {Phys. Rev. Lett.},
  volume = {113},
  issue = {9},
  pages = {093603},
  numpages = {5},
  year = {2014},
  month = {Aug},
  publisher = {American Physical Society},
  doi = {10.1103/PhysRevLett.113.093603},
  url = {https://link.aps.org/doi/10.1103/PhysRevLett.113.093603}
}

@article{uppu2020scalable,
author = {Ravitej Uppu  and Freja T. Pedersen  and Ying Wang  and Cecilie T. Olesen  and Camille Papon  and Xiaoyan Zhou  and Leonardo Midolo  and Sven Scholz  and Andreas D. Wieck  and Arne Ludwig  and Peter Lodahl },
title = {Scalable integrated single-photon source},
journal = {Science Advances},
volume = {6},
number = {50},
pages = {eabc8268},
year = {2020},
doi = {10.1126/sciadv.abc8268},
URL = {https://www.science.org/doi/abs/10.1126/sciadv.abc8268},
eprint = {https://www.science.org/doi/pdf/10.1126/sciadv.abc8268},
}

@article{lodahl2015interfacing,
  title = {Interfacing single photons and single quantum dots with photonic nanostructures},
  author = {Lodahl, Peter and Mahmoodian, Sahand and Stobbe, S\o{}ren},
  journal = {Rev. Mod. Phys.},
  volume = {87},
  issue = {2},
  pages = {347--400},
  numpages = {54},
  year = {2015},
  month = {May},
  publisher = {American Physical Society},
  doi = {10.1103/RevModPhys.87.347},
  url = {https://link.aps.org/doi/10.1103/RevModPhys.87.347}
}

@article{liu2024violation,
	author = {Liu, Shikai and Sandberg, Oliver August Dall'Alba and Chan, Ming Lai and Schrinski, Bj{\"o}rn and Anyfantaki, Yiouli and Nielsen, Rasmus B. and Larsen, Robert G. and Skalkin, Andrei and Wang, Ying and Midolo, Leonardo and Scholz, Sven and Wieck, Andreas D. and Ludwig, Arne and S{\o}rensen, Anders S. and Tiranov, Alexey and Lodahl, Peter},
	journal = {Nature Physics},
	number = {9},
	pages = {1429--1433},
	title = {Violation of Bell inequality by photon scattering on a two-level emitter},
	volume = {20},
	year = {2024}}

@article{weinfurter_experimental_1994,
	title = {Experimental {Bell}-{State} {Analysis}},
	volume = {25},
	issn = {0295-5075},
	url = {https://doi.org/10.1209/0295-5075/25/8/001},
	doi = {10.1209/0295-5075/25/8/001},
	number = {8},
	urldate = {2025-10-16},
	journal = {Europhysics Letters},
	author = {Weinfurter, H.},
	month = mar,
	year = {1994},
	pages = {559},
}

@misc{ostmann2025nonlinear,
      title={Nonlinear photonic architecture for fault-tolerant quantum computing}, 
      author={Maike Ostmann and Joshua Nunn and Alex E. Jones},
      year={2025},
      eprint={2510.06890},
      archivePrefix={arXiv},
      primaryClass={quant-ph},
      url={https://arxiv.org/abs/2510.06890}, 
}

@article{pedersen2020near,
  title={Near transform-limited quantum dot linewidths in a broadband photonic crystal waveguide},
  author={Pedersen, Freja T and Wang, Ying and Olesen, Cecilie T and Scholz, Sven and Wieck, Andreas D and Ludwig, Arne and LObl, Matthias C and Warburton, Richard J and Midolo, Leonardo and Uppu, Ravitej and others},
  journal={ACS Photonics},
  volume={7},
  number={9},
  pages={2343--2349},
  year={2020},
  publisher={ACS Publications}
}

@article{sangouard_quantum_2011,
    title = {Quantum repeaters based on atomic ensembles and linear optics},
    volume = {83},
    url = {https://link.aps.org/doi/10.1103/RevModPhys.83.33},
    doi = {10.1103/RevModPhys.83.33},
    number = {1},
    urldate = {2026-01-19},
    journal = {Reviews of Modern Physics},
    author = {Sangouard, Nicolas and Simon, Christoph and de Riedmatten, Hugues and Gisin, Nicolas},
    month = mar,
    year = {2011},
    pages = {33--80},
}

@article{scarani_security_2009,
    title = {The security of practical quantum key distribution},
    volume = {81},
    url = {https://link.aps.org/doi/10.1103/RevModPhys.81.1301},
    doi = {10.1103/RevModPhys.81.1301},
    number = {3},
    urldate = {2026-01-19},
    journal = {Reviews of Modern Physics},
    author = {Scarani, Valerio and Bechmann-Pasquinucci, Helle and Cerf, Nicolas J. and Dušek, Miloslav and Lütkenhaus, Norbert and Peev, Momtchil},
    month = sep,
    year = {2009},
    pages = {1301--1350},
}

@article{bennett_mixed-state_1996,
    title = {Mixed-state entanglement and quantum error correction},
    volume = {54},
    url = {https://link.aps.org/doi/10.1103/PhysRevA.54.3824},
    doi = {10.1103/PhysRevA.54.3824},
    number = {5},
    urldate = {2026-01-19},
    journal = {Physical Review A},
    author = {Bennett, Charles H. and DiVincenzo, David P. and Smolin, John A. and Wootters, William K.},
    month = nov,
    year = {1996},
    pages = {3824--3851},
}

@article{duan_long-distance_2001,
    title = {Long-distance quantum communication with atomic ensembles and linear optics},
    volume = {414},
    copyright = {2001 Macmillan Magazines Ltd.},
    issn = {1476-4687},
    url = {https://www.nature.com/articles/35106500},
    doi = {10.1038/35106500},
    number = {6862},
    urldate = {2026-01-19},
    journal = {Nature},
    author = {Duan, L.-M. and Lukin, M. D. and Cirac, J. I. and Zoller, P.},
    month = nov,
    year = {2001},
    pages = {413--418},
}

@article{chan2025tailoring,
  title = {Tailoring Fusion-Based Photonic Quantum Computing Schemes to Quantum Emitters},
  author = {Chan, Ming Lai and Bell, Thomas J. and Pettersson, Love A. and Chen, Susan X. and Yard, Patrick and S\o{}rensen, Anders S. and Paesani, Stefano},
  journal = {PRX Quantum},
  volume = {6},
  issue = {2},
  pages = {020304},
  numpages = {18},
  year = {2025},
  month = {Apr},
  publisher = {American Physical Society},
  doi = {10.1103/PRXQuantum.6.020304},
  url = {https://link.aps.org/doi/10.1103/PRXQuantum.6.020304}
}

@article{chang_quantum_2014,
    title = {Quantum nonlinear optics — photon by photon},
    volume = {8},
    copyright = {2014 Springer Nature Limited},
    issn = {1749-4893},
    url = {https://www.nature.com/articles/nphoton.2014.192},
    doi = {10.1038/nphoton.2014.192},
    number = {9},
    urldate = {2026-04-01},
    journal = {Nature Photonics},
    publisher = {Nature Publishing Group},
    author = {Chang, Darrick E. and Vuletić, Vladan and Lukin, Mikhail D.},
    month = sep,
    year = {2014},
    pages = {685--694},
}

@article{calsamiglia_maximum_2001,
    title = {Maximum efficiency of a linear-optical {Bell}-state analyzer},
    volume = {72},
    issn = {1432-0649},
    url = {https://doi.org/10.1007/s003400000484},
    doi = {10.1007/s003400000484},
    number = {1},
    urldate = {2026-04-01},
    journal = {Applied Physics B},
    author = {Calsamiglia, J. and Lütkenhaus, N.},
    month = jan,
    year = {2001},
    pages = {67--71},
}

@article{braunstein_measurement_1995,
    title = {Measurement of the {Bell} operator and quantum teleportation},
    volume = {51},
    url = {https://link.aps.org/doi/10.1103/PhysRevA.51.R1727},
    doi = {10.1103/PhysRevA.51.R1727},
    number = {3},
    urldate = {2026-04-01},
    journal = {Physical Review A},
    publisher = {American Physical Society},
    author = {Braunstein, Samuel L. and Mann, A.},
    month = mar,
    year = {1995},
    pages = {R1727--R1730},
}

@misc{bartolucci_comparison_2025,
    title = {Comparison of schemes for highly loss tolerant photonic fusion based quantum computing},
    url = {https://arxiv.org/abs/2506.11975v1},
    urldate = {2026-04-02},
    journal = {arXiv.org},
    author = {Bartolucci, Sara and Bell, Tom and Bombin, Hector and Birchall, Patrick and Bulmer, Jacob and Dawson, Christopher and Farrelly, Terry and Gartenstein, Samuel and Gimeno-Segovia, Mercedes and Litinski, Daniel and Liu, Yehua and Knegjens, Robert and Nickerson, Naomi and Olivo, Andrea and Pant, Mihir and Patil, Ashlesha and Roberts, Sam and Rudolph, Terry and Sparrow, Chris and Tuckett, David and Veitia, Andrzej},
    month = jun,
    year = {2025},
}

@article{birnbaum_photon_2005,
    title = {Photon blockade in an optical cavity with one trapped atom},
    volume = {436},
    copyright = {2005 Springer Nature Limited},
    issn = {1476-4687},
    url = {https://www.nature.com/articles/nature03804},
    doi = {10.1038/nature03804},
    number = {7047},
    urldate = {2026-04-01},
    journal = {Nature},
    publisher = {Nature Publishing Group},
    author = {Birnbaum, K. M. and Boca, A. and Miller, R. and Boozer, A. D. and Northup, T. E. and Kimble, H. J.},
    month = jul,
    year = {2005},
    pages = {87--90},
}

@article{hacker_photonphoton_2016,
    title = {A photon–photon quantum gate based on a single atom in an optical resonator},
    volume = {536},
    copyright = {2016 Springer Nature Limited},
    issn = {1476-4687},
    url = {https://www.nature.com/articles/nature18592},
    doi = {10.1038/nature18592},
    number = {7615},
    urldate = {2026-04-01},
    journal = {Nature},
    publisher = {Nature Publishing Group},
    author = {Hacker, Bastian and Welte, Stephan and Rempe, Gerhard and Ritter, Stephan},
    month = aug,
    year = {2016},
    pages = {193--196},
}

@article{tiarks_optical_2016,
    title = {Optical {$\pi$} phase shift created with a single-photon pulse},
    volume = {2},
    url = {https://www.science.org/doi/10.1126/sciadv.1600036},
    doi = {10.1126/sciadv.1600036},
    number = {4},
    urldate = {2026-04-02},
    journal = {Science Advances},
    publisher = {American Association for the Advancement of Science},
    author = {Tiarks, Daniel and Schmidt, Steffen and Rempe, Gerhard and Dürr, Stephan},
    month = apr,
    year = {2016},
    pages = {e1600036},
}

@article{tiarks_photonphoton_2019,
    title = {A photon–photon quantum gate based on {Rydberg} interactions},
    volume = {15},
    copyright = {2018 The Author(s), under exclusive licence to Springer Nature Limited},
    issn = {1745-2481},
    url = {https://www.nature.com/articles/s41567-018-0313-7},
    doi = {10.1038/s41567-018-0313-7},
    number = {2},
    urldate = {2026-04-01},
    journal = {Nature Physics},
    publisher = {Nature Publishing Group},
    author = {Tiarks, Daniel and Schmidt-Eberle, Steffen and Stolz, Thomas and Rempe, Gerhard and Dürr, Stephan},
    month = feb,
    year = {2019},
    pages = {124--126},
}

@article{sparrow2018simulating,
  title={Simulating the vibrational quantum dynamics of molecules using photonics},
  author={Sparrow, Chris and Mart{\'\i}n-L{\'o}pez, Enrique and Maraviglia, Nicola and Neville, Alex and Harrold, Christopher and Carolan, Jacques and Joglekar, Yogesh N and Hashimoto, Toshikazu and Matsuda, Nobuyuki and O’Brien, Jeremy L and others},
  journal={Nature},
  volume={557},
  number={7707},
  pages={660--667},
  year={2018},
  publisher={Nature Publishing Group UK London}
}

@article{Tomm2023,
  title        = {Photon bound state dynamics from a single artificial atom},
  author       = {Tomm, Natasha and Mahmoodian, Sahand and Antoniadis, Nadia O. and Schott, R\"udiger and Valentin, Sascha R. and Wieck, Andreas D. and Ludwig, Arne and Javadi, Alisa and Warburton, Richard J.},
  journal      = {Nature Physics},
  volume       = {19},
  pages        = {857--862},
  year         = {2023},
  doi          = {10.1038/s41567-023-01997-6},
  publisher    = {Nature Publishing Group}
}

@misc{dessertaine_enhanced_2024,
    title = {Enhanced {Fault}-tolerance in {Photonic} {Quantum} {Computing}: {Comparing} the {Honeycomb} {Floquet} {Code} and the {Surface} {Code} in {Tailored} {Architecture}},
    shorttitle = {Enhanced {Fault}-tolerance in {Photonic} {Quantum} {Computing}},
    url = {https://arxiv.org/abs/2410.07065v3},
    urldate = {2026-04-07},
    journal = {arXiv.org},
    author = {Dessertaine, Théo and Bourdoncle, Boris and Denys, Aurélie and de Gliniasty, Grégoire and d'Istria, Pierre Colonna and Valentí-Rojas, Gerard and Mansfield, Shane and Hilaire, Paul},
    month = oct,
    year = {2024},
}

@article{dreesen_suppressing_2018,
    title = {Suppressing phonon decoherence of high performance single-photon sources in nanophotonic waveguides},
    volume = {4},
    issn = {2058-9565},
    url = {https://doi.org/10.1088/2058-9565/aadbb8},
    doi = {10.1088/2058-9565/aadbb8},
    number = {1},
    urldate = {2026-01-13},
    journal = {Quantum Science and Technology},
    publisher = {IOP Publishing},
    author = {Dreeßen, C L and Ouellet-Plamondon, C and Tighineanu, P and Zhou, X and Midolo, L and Sørensen, A S and Lodahl, P},
    month = sep,
    year = {2018},
    pages = {015003},
}

@article{kuhlmann_charge_2013,
    title = {Charge noise and spin noise in a semiconductor quantum device},
    volume = {9},
    copyright = {2013 Springer Nature Limited},
    issn = {1745-2481},
    url = {https://www.nature.com/articles/nphys2688},
    doi = {10.1038/nphys2688},
    number = {9},
    urldate = {2024-11-20},
    journal = {Nature Physics},
    publisher = {Nature Publishing Group},
    author = {Kuhlmann, Andreas V. and Houel, Julien and Ludwig, Arne and Greuter, Lukas and Reuter, Dirk and Wieck, Andreas D. and Poggio, Martino and Warburton, Richard J.},
    month = sep,
    year = {2013},
    pages = {570--575},
}

@article{molmer1993monte,
author = {Klaus M{\o}lmer and Yvan Castin and Jean Dalibard},
journal = {J. Opt. Soc. Am. B},
number = {3},
pages = {524--538},
publisher = {Optica Publishing Group},
title = {Monte Carlo wave-function method in quantum optics},
volume = {10},
month = {Mar},
year = {1993},
url = {https://opg.optica.org/josab/abstract.cfm?URI=josab-10-3-524},
doi = {10.1364/JOSAB.10.000524},
}

@article{witthaut_photon_2010,
    title = {Photon scattering by a three-level emitter in a one-dimensional waveguide},
    volume = {12},
    issn = {1367-2630},
    url = {https://doi.org/10.1088/1367-2630/12/4/043052},
    doi = {10.1088/1367-2630/12/4/043052},
    number = {4},
    urldate = {2026-04-08},
    journal = {New Journal of Physics},
    author = {Witthaut, D and Sørensen, A S},
    month = apr,
    year = {2010},
    pages = {043052},
}


\newpage 
\onecolumngrid
\clearpage

\pagenumbering{arabic}
\appendix
\renewcommand{\thesection}
{\Alph{section}}
\renewcommand{\thefigure}{\thesection\arabic{figure}}
\setcounter{figure}{0} 
\renewcommand{\thetable}{\thesection\arabic{table}}
\setcounter{table}{0} 


Contents 
\begin{enumerate}
    \item Appendix \ref{appendix:expedetails}: Experimental setup and optical characterisation
    \item Appendix \ref{appendix:Data_analysis}: Mapping photon statistics to sorting probability
    \item Appendix \ref{appendix:theory}: Theoretical model of photon scattering with noise
    \item Appendix \ref{appendix:filtering}: Temporal Filtering
    \item Appendix \ref{sec:single_vs_chiral}: Single-sided, double-sided and chiral Waveguides 
    \item Appendix \ref{appendix:RTsorter}: Comparison to single emitter sorter architectures
    \item Appendix \ref{appendix:fusion}: Application: FBQC with photon-sorter-boosted fusions
    \item Appendix \ref{appendix:qkd}: Application:  Quantum key distribution using photon sorter
\end{enumerate}

\newpage
\section{Experimental setup and optical characterisation}
\label{appendix:expedetails}
\subsection{Experiment Configuration}
\begin{figure}[ht]
    \centering
    \includegraphics[width=0.8\textwidth]{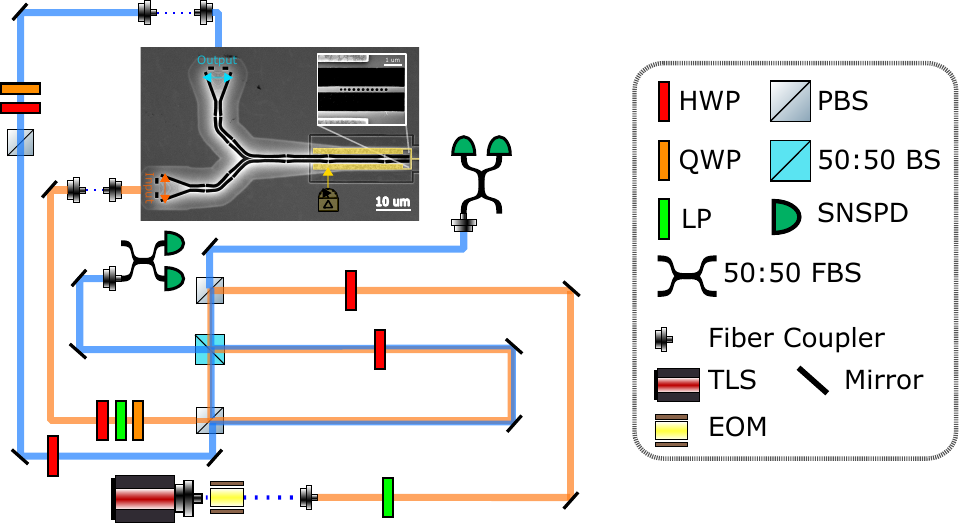}
    \caption{Sketch of the experimental setup. A free-space, self-stabilised double-path interferometer is modularly integrated with the non-linear optical circuit, forming the photonic sorter (shown in main text Fig. 1a). The \textbf{orange line} highlights the path for preparing input states, which are then coupled into the non-linear device. The \textbf{blue line} indicates the path used to facilitate the photon sorting protocols.}
    \label{fig:suppl:setup}
\end{figure}

As shown in Fig.~\ref{fig:suppl:setup} the non-linear device consisting of QDs two-level system integrated in a nanobeam waveguide is  incorporated within an unbalanced Mach-Zehnder Interferometer. This configuration creates a photonic circuit capable of separating photons based on their number state. We verified the performance of this circuit and its protocols by inputting weak coherent state ($\ket{\alpha}$) pulses containing up to two photons. We use a compact, double-pass interferometer to cancel phase fluctuations.

The \textbf{orange path} is used to prepare the time-bin encoded input photon states. A continuously wavelength tunable laser (CTL) is combined with an electro-optic modulator (EOM), which is driven by an arbitrary waveform generator (AWG), to generate Gaussian pulses of various durations between 700 ps and 1.2 ns. These pulse durations serve as a non-linear tuning knob.
The laser pulses propagate through a linear polariser (LP) and a half-wave plate (HWP), ensuring V polarisation, and therefore are reflected by the polarising beam splitter (PBS). 
A 50/50 beam splitter (BS) then splits the pulses into two paths — a short and a long arm - the latter including an additional HWP. This configuration generates time-bin pulses corresponding to the early state $\ket{e}$ and the late state $\ket{l}$ upon recombination at the PBS.
A quarter-wave plate (QWP) converts the V-polarised early pulse ($\ket{e}$) and H-polarised late pulse ($\ket{l}$) into left- and right-circular polarisations, respectively, producing equal amplitudes after passing through the \textit{LP}, which is used to set the relative phase between the two time-bins. 
Finally, a HWP is used to maximise coupling efficiency into the optical fiber that delivers the excitation photons to the nanophotonic device (SEM image in Fig.~\ref{fig:suppl:setup}), which is mounted in a cryostat maintained at 1.6 K.

The \textbf{blue path} directs the photons back into the same interferometer after the nonlinear interaction with the QD two-level system. The PBS equally splits the out-coupled photons into the short and long arms. Interference occurs at the BS when the early photons are reflected into the long arm, while the late photons are transmitted through the short arm. Although this configuration introduces a 50 \% loss, it does not affect the photon-sorting fidelity. Such loss can be readily mitigated by active switching employing a resonant electro-optic modulator (EOM). A pseudo photon-number measurement is implemented by inserting a 50/50 fiber BS whose two output ports are connected to superconducting nanowire single-photon detectors (SNSPDs).
A detailed characterisation of the interferometer is provided in Ref.~\cite{nielsen_programmable_2025}.

\subsection{Nanophotonic Device and Optical Characterisation} 
\begin{figure}[ht]
    \centering
    \includegraphics{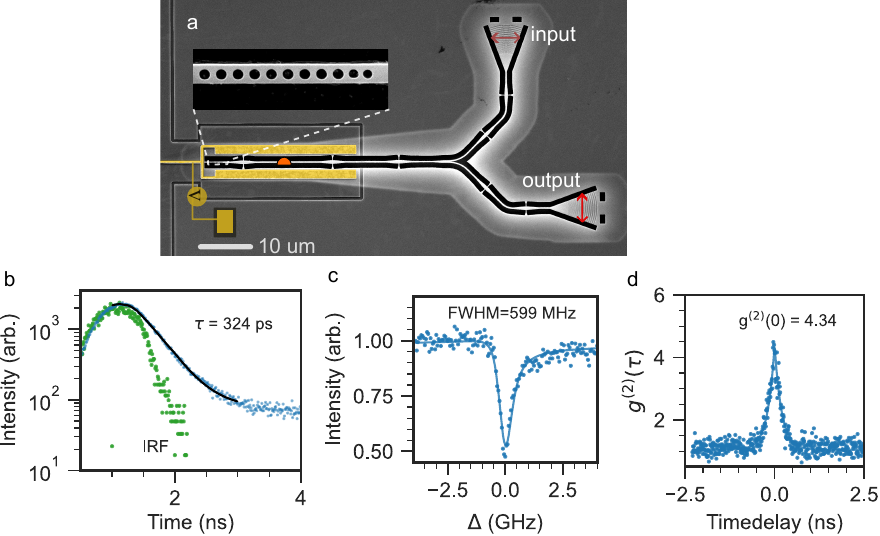}
    \caption{
    (a) SEM image of the device. The photonic crystal mirrors are adiabatically interfaced with a nanobeam waveguide. The Y-splitter terminates with a vertically polarised shallow-etched grating coupler for the input and a horizontally polarised coupler for the output.
    (b) Lifetime measurement of the selected QD. The green dots represent the instrument response function, while the light-blue trace shows the QD decay curve fitted with a single-exponential model (black line).
    (c) Measured output spectrum as a function of detuning ($\Delta$) of the QD from its resonant frequency, showing a Fano-like transmission profile.
    (d) Second-order correlation function $g^{(2)}(\tau)$ recorded at zero detuning ($\Delta = 0$).
    }
    \label{fig:suppl:1_sided NB}
\end{figure}

The device in this study is integrated into a nanobeam waveguide shown in Fig.~\ref{fig:suppl:1_sided NB}(a). One side of the waveguide is terminated with a photonic mirror that exhibits high reflectivity and a broad bandwidth. The opposite termination integrates an on-chip Y-beamsplitter that interfaces with shallow-etched grating couplers, enabling efficient coupling of light into and out of the photonic chip. The implementation of such an on-chip beam splitter, which enables spatial separation between the input and output channels, significantly improves experimental robustness and reduces experimental complexity by eliminating the need for cross-polarisation. In practice when using weak coherent states as inputs, an unbalanced beam splitter (e.g., 99:1) is preferred over a 50:50 Y-splitter configuration and can be readily implemented within a compact footprint using evanescently coupled waveguides. For applications requiring high in- and out-coupling efficiencies, using a Faraday mirror configuration instead of an on-chip beam splitter to separate the incoming and outgoing beams would be beneficial.
A locally defined metal gate, highlighted in yellow, enables the application of an external bias to suppress charge noise, thereby reducing spectral wandering and providing a tunable control parameter for the QD–light interaction. The localised contact not only improves fabrication yield but also enhances the device response time to a few MHz, as reported in Ref.~\cite{pedersen2020near}.

The QD decay rate was characterised using time-resolved first-order correlation measurements $g^{(1)}(t)$ (Fig.~\ref{fig:suppl:1_sided NB}), revealing a lifetime of $1/\gamma = 324$ ps, which corresponds to a natural linewidth of $\gamma / 2\pi = 491$ MHz. A weak coherent laser with an average photon number $\langle n \rangle < 0.1$ was tuned to the QD resonance frequency, while the external bias was scanned across the resonant voltage. The resulting output spectrum, shown in Fig.~\ref{fig:suppl:1_sided NB}c, exhibits a Fano-like transmission profile. This behaviour arises from mirror-induced phase differences between the left- and right-propagating fields, as well as small residual reflections from the Y-splitter, leading to a complex interference pattern. The second-order correlation function $g^{(2)}(\tau)$, shown in Fig.~\ref{fig:suppl:1_sided NB}d, exhibits a pronounced bunching peak at zero time delay, with $g^{(2)}(0) = 4.34$, limited primarily by residual laser background. This observation indicates that the light–QD nonlinear interaction redistributes the photon statistics, forming the basis for the photon-sorting mechanism. The characterisations here allow us to extract the relevant QD parameters (such as the lifetime and $\beta$-factor) and inform our choice of QD to use in the experiment.

\section{Map photon statistics to sorting probability}
\label{appendix:Data_analysis}
\begin{figure}[]
    \centering
    \includegraphics{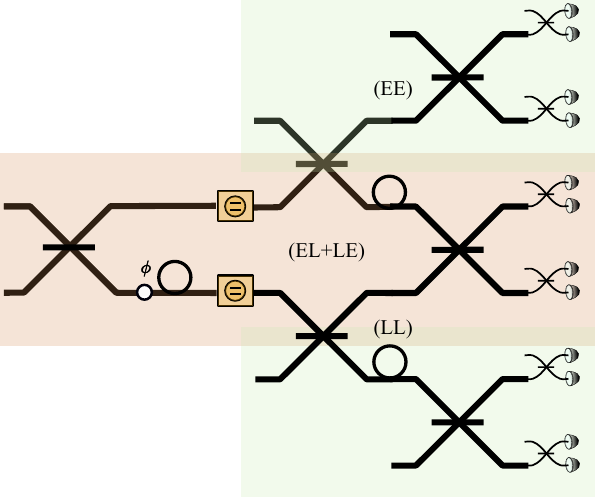}
    \caption{The time-bin encoding of the photons can be converted into distinct spatial modes. This time-to-space projection provides a straightforward way to describe the photon statistics measured at the output ports.}
    \label{fig:S_psExperiment}
\end{figure}

The self-stabilising time-bin interferometer generates three time bins before the final detection. These are: early-early (EE), early-late + late-early (EL+LE), and late-late (LL). Ideally, we want all clicks to occur in the EL+LE bin, where the two modes interfere (which can be implemented using active switching, as discussed in the main text), but we can use the other bins (EE and LL) for normalisation. A schematic of the full experiment, where the temporal modes are drawn explicitly as spatial modes, is shown in Fig.~\ref{fig:S_psExperiment}.
\subsection{One-photon statistics}
Assuming a single-photon input state, the number of counts in the LL time bin in the first mode is given by
\begin{equation}
    C_\text{LL,1}=\frac{1}{2}\frac{1}{2}\frac{1}{2}\eta_\text{L,1}\eta_\text{L,2}\eta_\text{D1},
\end{equation}
where we assume a fixed integration time and repetition rate, which we factor out since they are constants across experiments.
The factors of $\frac{1}{2}$ come from the probability of a photon going to the short arm at the first pass, going to the short arm at the second pass, and then going to the right detector. All the beam splitters associated with these transformations are assumed to be 50:50. $\eta_{L,1}$ ($\eta_{L,2}$) denotes the efficiency of the first (second) pass of the late arm of the interferometer and $\eta_{D,1}$ the efficiency of detector 1.
 We assume that the efficiency of coupling to and from the chip is independent of the time bin; thus, this factor cancels out across all terms. In principle, there is also a factor of $\frac{1}{2}$ from the fact that each mode passes through one out of two detectors after a fiber-based 50:50 BS. Since every photon experiences this, it cancels out in the end. A similar argument can be made for detection in the EE timebin, where the efficiencies are now of the short path for the first and second pass.
The number of counts in the central bin (EL + LE) is given by a similar argument. 
\begin{equation}
    C_\text{C,1}=\frac{1}{2}P_\text{C,1}(\eta_\text{L,1}\eta_\text{E,2}+\eta_\text{E,1}\eta_\text{L,2})\eta_\text{D1},
\end{equation}
where $P_\text{C,1}$ is the lossless probability for a photon to click in the central mode, which depends on the linear phase and nonlinearity. The factor of $\frac{1}{2}$ is due to the probabilistic nature that a photon that went early on the first pass has to go late on the second, or vice versa.
We are interested in the value $P_\text{C,1}$. In the experiment, we ensure that the early and late path through the interferometer has the same efficiency, which means that $\eta_\text{L,1}=\eta_\text{E,1}$ and $\eta_\text{L,2}=\eta_\text{E,2}$. We can then extract  $P_\text{C,1}$ from 
\begin{equation}
    \frac{C_\text{C,1}}{\sqrt{C_\text{LL,1}}\sqrt{C_\text{EE,1}}}=2^3P_\text{C,1}.
\end{equation}
Suppose we normalise this probability to the probability of getting clicks in the other mode $P_\text{C,2}$. In that case, the factor of 8 will cancel, and we can estimate the one-photon probability for both interferometer modes.
\begin{equation}
    \mathcal{P}_{1}=\frac{P_\text{C,1}}{P_\text{C,1}+P_\text{C,2}}
\end{equation}
\subsection{Two-photon statistics}
To find the two-photon statistics, we can apply a similar analysis to that used for the one-photon statistics. In this case, we will examine the off-diagonal side peaks. The number of coincidences in the (EE/EE) temporal bin is 
\begin{equation}
    C_\text{EE/EE,20}=\frac{1}{2^2}\frac{1}{2^2}\frac{1}{2^2}\frac{1}{2}\eta_1^2\eta_2^2\eta_\text{D1,a}\eta_\text{D1,b}.
\end{equation}
The factors of 2 arise from both photons going to the early time-bin for the first pass and the second pass of the interferometer. Both photons are sent to mode 1 for detection, and then they are split up for pseudo-number-resolving detection. For efficiency, we again assume that the early and late paths are equally coupled in the first and second passes, respectively. A similar analysis can be made for the LL/LL time bin.
The number of coincidences in the central time bin is given by 
\begin{equation}
    C_\text{C/C,20} = \frac{1}{2}P_\text{C/C,20}\eta_1^2\eta_2^2\eta_\text{D1,a}\eta_\text{D1,b},
\end{equation}
where the factor of $\frac{1}{2}$ is due to the photons having to split up at the pseudo-number-resolving detectors.
The expressions for bunching in the other mode (02) are effectively equivalent to these, but the expressions for the antibunching case (11) are given by
\begin{equation}
    \begin{split}
        C_\text{EE/EE,11}=\frac{1}{2^2}\frac{1}{2^2}\frac{1}{2^2}\frac{1}{2}\eta_1^2\eta_2^2(\eta_\text{D1,a}\eta_\text{D2,a}+\eta_\text{D1,a}\eta_\text{D2,b}+\eta_\text{D1,b}\eta_\text{D2,a}+\eta_\text{D1,b}\eta_\text{D2,b}),\\
        C_\text{C/C,11}=\frac{1}{4} P_\text{C/C,11}\eta_1^2\eta_2^2(\eta_\text{D1,a}\eta_\text{D2,a}+\eta_\text{D1,a}\eta_\text{D2,b}+\eta_\text{D1,b}\eta_\text{D2,a}+\eta_\text{D1,b}\eta_\text{D2,b}),
    \end{split}
\end{equation}
where the factor of $\frac{1}{4}$ in the central bin is due to the fact that there are 4 different detection combinations.
The normalised probabilities for the different detector combinations are given by
\begin{equation}
    \mathcal{P}_{20}=\frac{P_\text{C/C,20}}{P_\text{C/C,20}+P_\text{C/C,02}+P_\text{C/C,11}}
\end{equation}

\section{Theoretical model of photon scattering with noise}
~\label{appendix:theory}
Here we give the details of the theoretical model. First in the noiseless case and later in the presence of pure dephasing and spectral diffusion.

\subsection{Noiseless scattering}
\begin{figure}
    \centering
    \includegraphics[width=1\linewidth]{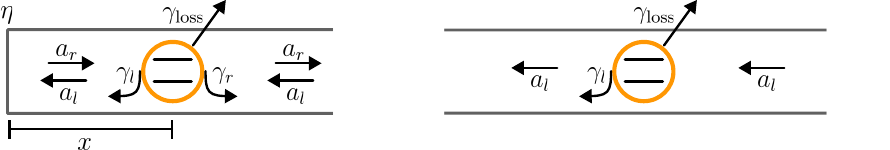}
    \caption{Single-sided waveguide with one end terminated by a  mirror in comparison to a chiral waveguide.}
    \label{fig:chiral_vs_single}
\end{figure}

The scattering of photons from a two-level system has been studied in great detail theoretically \cite{fan2010input,ramos2018correlated}. Here we present a recap of the most important results. A schematic of the system is shown in Fig.~\ref{fig:chiral_vs_single}, where the system is a single-sided waveguide with a mirror, but we model it as a chiral system. A justification for this model is given in section \ref{sec:single_vs_chiral} and also derived in \cite{witthaut_photon_2010, fan2010input}. The system can be described by the following Hamiltonian, which contains three terms 
\begin{equation}
    H = H_\text{qd}+H_\text{ph}+H_\text{qd-ph},
\end{equation}
with $H_\text{qd}$ describing the two-level system, $H_\text{ph}$ describing the propagating photons, and $H_\text{qd-ph}$ describing the interaction. In the noiseless case, the qubit Hamiltonian is given by
\begin{equation}
    H_\text{qd}=\frac{1}{2}\omega_0\sigma_z,
\end{equation}
with the diagonal Pauli operator  $\sigma_z=\ket{e}\bra{e}-\ket{g}\bra{g}$ where $\ket{e}$ ($\ket{g}$) denotes the excited (ground) state of the qubit and $\omega_0$ the angular frequency of the transition.
The Hamiltonian of the free propagating photons is given by
\begin{equation}
    H_\text{ph}=\int\dd\omega a_\omega^\dagger a_\omega + \int\dd\omega b_\omega^\dagger b_\omega,
\end{equation}
where $a_\omega$ destroys a guided photon of frequency $\omega$, and $b_\omega$ destroys a non-guided (lost) photon.
Finally, the Hamiltonian of the interaction under the rotating wave approximation is given by
\begin{equation}
    H_\text{qd-ph}=\sqrt{\frac{\gamma}{2\pi}}\int\dd\omega (\sigma_-a_\omega^\dagger + \sigma_+a_\omega )+\sqrt{\frac{\gamma_\text{loss}}{2\pi}} \int\dd\omega (\sigma_-b_\omega^\dagger + \sigma_+b_\omega ),
\end{equation}
where $\gamma$ is the emitter-waveguide mode coupling rate and $\gamma_\text{loss}$ is the emitter-unguided mode coupling rate. From these rates, we can define the overall waveguide coupling as $\beta=\gamma/(\gamma+\gamma_\text{loss})=\gamma/\Gamma$ with $\Gamma=\gamma+\gamma_\text{loss}$.
Following the standard procedure of input-output theory, it is possible to integrate out the photons and be left with the Heisenberg equations for the qubit operators
\begin{align}
    \frac{\dd\sigma^-}{\dd t}=& -\left(\frac{\Gamma}{2}+i\omega_0\right)\sigma^-+i\sigma_z\sqrt{\gamma}a_\text{in}(t)+i\sigma_z\sqrt{\gamma_\text{loss}}b_\text{in}(t),\\
    \frac{\dd\sigma_z}{\dd t} =& -\Gamma(\sigma_z+1)-2i\sqrt{\gamma}(\sigma^+a_\text{in}(t)-\sigma^-a_\text{in}^\dagger(t))-2i\sqrt{\gamma_\text{loss}}(\sigma^+b_\text{in}(t)-\sigma^-b_\text{in}^\dagger(t)).
\end{align}
The input field Heisenberg operators for the guided and unguided modes are given by a Fourier transform
\begin{align}
    a_\text{in}(t)=\frac{1}{\sqrt{2\pi}}\int\dd\omega e^{-i \omega(t-t_0)}a_\omega(t_0), \quad
    b_\text{in}(t)=\frac{1}{\sqrt{2\pi}}\int\dd\omega e^{-i \omega(t-t_0)}b_\omega(t_0).
\end{align}
The output field operators are given by the input-output relations as
\begin{equation}
    a_\text{out}(t)=a_\text{in}(t)-i\sqrt{\gamma}\sigma^-(t).
\end{equation}
The single-photon transmission coefficient is given by
\begin{equation}
    t_k=1-\frac{\gamma}{\Gamma/2-i(\omega-\omega_0)},
\end{equation}
where the Fourier transform of the excited state amplitude gives the excitation of the two-level system by a single-photon pulse
\begin{equation}
    s_k=-i\left(\frac{\sqrt{\gamma}}{\Gamma/2-i(\omega-\omega_0)}\right)
\end{equation}
with the relation between the transmission coefficient and excitation,
\begin{equation}
    t_k=1-i\sqrt{\gamma}s_k.
\end{equation}
The scattering matrix ($S$) element of an incident single-photon with frequency $k$ and output frequency $p$ is given by
\begin{equation}
    \bra{p}S\ket{p}=t_k\delta(k-p).
\end{equation}
The two-photon scattering matrix element of incident photons $k_1,k_2$ and output photons $p_1,p_2$ is given by
\begin{equation}
\begin{split}
    & \bra{0}a_\text{out}(p_1)a_\text{out}(p_2)a_\text{in}(k_1)^\dagger a_\text{in}(k_2)^\dagger \ket{0}=\\ 
    & \quad t_{k_1} t_{k_2} (\delta(k_2-p_1)\delta(k_1-p_2)+\delta(k_1-p_1)\delta(k_2-p_2))+
    i\frac{1}{\pi}\sqrt{\gamma}\delta(k_1+k_2-p_1-p_2)s_{p_1}s_{p_2}(s_{k_1}+s_{k_2}).
\end{split}
\end{equation}
This equation contains two terms, which are typically denoted as the elastic and inelastic scattering terms. The elastic term is twice the one-photon scattering matrix, where the input and output photons pairwise exchange energy. The inelastic term contrains combined energy conservation of the two input and two output photons, allowing them to exchange energy with each other and \textit{interact}. 
%

%
\subsection{Pure dephasing}
The resonance frequency of the quantum emitter fluctuates over time due to different noise processes. Phonons coupled to the emitter can cause frequency shifts in the $\sim 10$ ps time scale. This process also causes partial distinguishability in a single-photon source from a quantum emitter \cite{dreesen_suppressing_2018}. This noise will affect neighbouring scatterings in an experiment and cause the first and second scattering to differ. In this experiment, this results in the one-photon states being output in the wrong mode because the superposition no longer exhibits perfect interference. It also causes the events where the two photon end up in different arms for the two-photon probabilities. To model this effect, we can modify the qubit Hamiltonian with a stochastic variable $\Delta(t)$ \cite{ramos2018correlated}
\begin{equation}
    H_\text{qd}=\frac{1}{2}[\omega_0+\Delta(t)]\sigma_z,
\end{equation}
where we assume the stochastic average is zero $\ev{\ev{\Delta(t)}}=0$. This means the equations of motion become the stochastic Heisenberg-Langevin equations \cite{ramos2018correlated}
\begin{align}
    \frac{\dd\sigma^-}{\dd t}=& -\left(\frac{\Gamma}{2}+i[\omega_0+\Delta(t)]\right)\sigma^-+i\sigma_z\sqrt{\gamma}a_\text{in}(t)+i\sigma_z\sqrt{\gamma_\text{loss}}b_\text{in}(t),\\
    \frac{\dd\sigma_z}{\dd t} =& -\Gamma(\sigma_z+1)-2i\sqrt{\gamma}(\sigma^+a_\text{in}(t)-\sigma^-a_\text{in}^\dagger(t))-2i\sqrt{\gamma_\text{loss}}(\sigma^+b_\text{in}(t)-\sigma^-b_\text{in}^\dagger(t)).
\end{align}
If we assume an uncorrelated Markovian white noise, which has the autocorrelation $\langle \Delta(0) \Delta(\tau) \rangle = 2\gamma_d \delta(\tau)$ with $\gamma_d$ the pure dephasing rate then the average transmission coefficient becomes \cite{ramos2018correlated},
\begin{equation}
    \ev{\ev{t_k}}=1-\frac{\gamma}{\Gamma/2+\gamma_d-i(\omega-\omega_0)}.
\end{equation}
This means that this uncorrelated noise corresponds to replacing $\Gamma/2 \rightarrow \Gamma/2 + \gamma_d$, similar to the standard description of pure dephasing. Indeed this model can be shown to be equivalent to the standard description of pure dephasing with dephasing operators in a master equation. Details of how to handle correlated noise contributions are found in Ref. \cite{ramos2018correlated}.
To model the experiment fully, we employ a jump-operator approach resembling the Monte-carlo wavefunction approach \cite{molmer1993monte}. Using the transmission coefficient above will result in unnormalised output states even for unit waveguide coupling $\beta=1$ with $\Tr(\rho)<1$ if $\rho$ is the output state from a scattering event using the \textit{noisy transmission coefficient} since these only describe the coherent (no-jump) evolution and neglect he incoherent contributions (the jump contribution). In order to normalise the states, we assume that the remainder is contributed to a pure dephasing decay, with a probability given by $P_{\gamma_d}$. In this model, we assume that photons emitted due to pure dephasing are completely distinguishable (orthogonal quantum states) from the photons emitted via the regular decay channel $\gamma$. We also assume photons from the pure dephasing decay channel are distinguishable from each other. The combined state at the output is thus
\begin{equation}
    \rho=\rho+\rho_{\gamma_d},
\end{equation}
where $\rho_{\gamma_d}=\sum_i^\infty \rho_{\gamma_d,i}$ with any $\rho_{\gamma_d,i}$ being distinguishable from one another. 
Note that in the case where $\beta<1$, the output states will still not be normalised due to photon losses to unguided modes, which we effectively trace out by only considering the guided photons. We assume that the ratio of total decay to pure dephasing remains the same when changing the coupling rate $\beta$. Thus, we can use the case of $\beta=1$ to estimate the dephasing rate (relative to the total decay) for cases with $\beta<1$.

\subsection{Spectral diffusion}
Another type of noise is slow charge noise \cite{kuhlmann_charge_2013}, which can cause drift on the $\sim \mu s$ timescale. This will not impact neighbouring scattering events, which in our case are separated by 5 ns, as this process does not influence the HOM visibility between photons emitted hundreds of nanoseconds apart \cite{uppu2020scalable}. It does, however, reduce the accuracy with which we can apply a resonant laser for scattering when running experiments that run for hours. We model this noise by sampling the frequency from a Gaussian distribution, where the standard deviation of the distribution dictates the amount of spectral diffusion. We find that experiments that integrate in the hour time-scale compared to a few minutes experiences slightly higher spectral diffusion, which was also reported in Ref.~\cite{liu2024violation}. The effect can  be modelled by convoluting frequency-dependent values (like photon statistics) with a Gaussian function of the same width.
\begin{equation}
    \text{SD}(\omega_\text{SD})=\frac{1}{\sqrt{2\pi\sigma_\text{SD}^2}}e^{-(\omega_0-\omega_\text{SD}/(2\sigma_{SD}^2))}
\end{equation}

\subsection{Fitting}
Using the methods described above we can solve the scattering dynamics with the noise terms from finite waveguide coupling, pure dephasing and spectral diffusion. In experiment we naturally post-select lost photons due to waveguide coupling away (i.e. they are not measured since they are lost), and we assume pure dephasing leads to perfectly distinguishable photons. The output density matrix after scattering is propagated through the second beam splitter before being measured with pseudo-number-resolving detectors. We perform a joint fitting of the one- and two-photon statistics.

\section{Temporal filtering} \label{appendix:filtering}

\begin{figure}[h]
    \centering
    \includegraphics[width=0.8\linewidth]{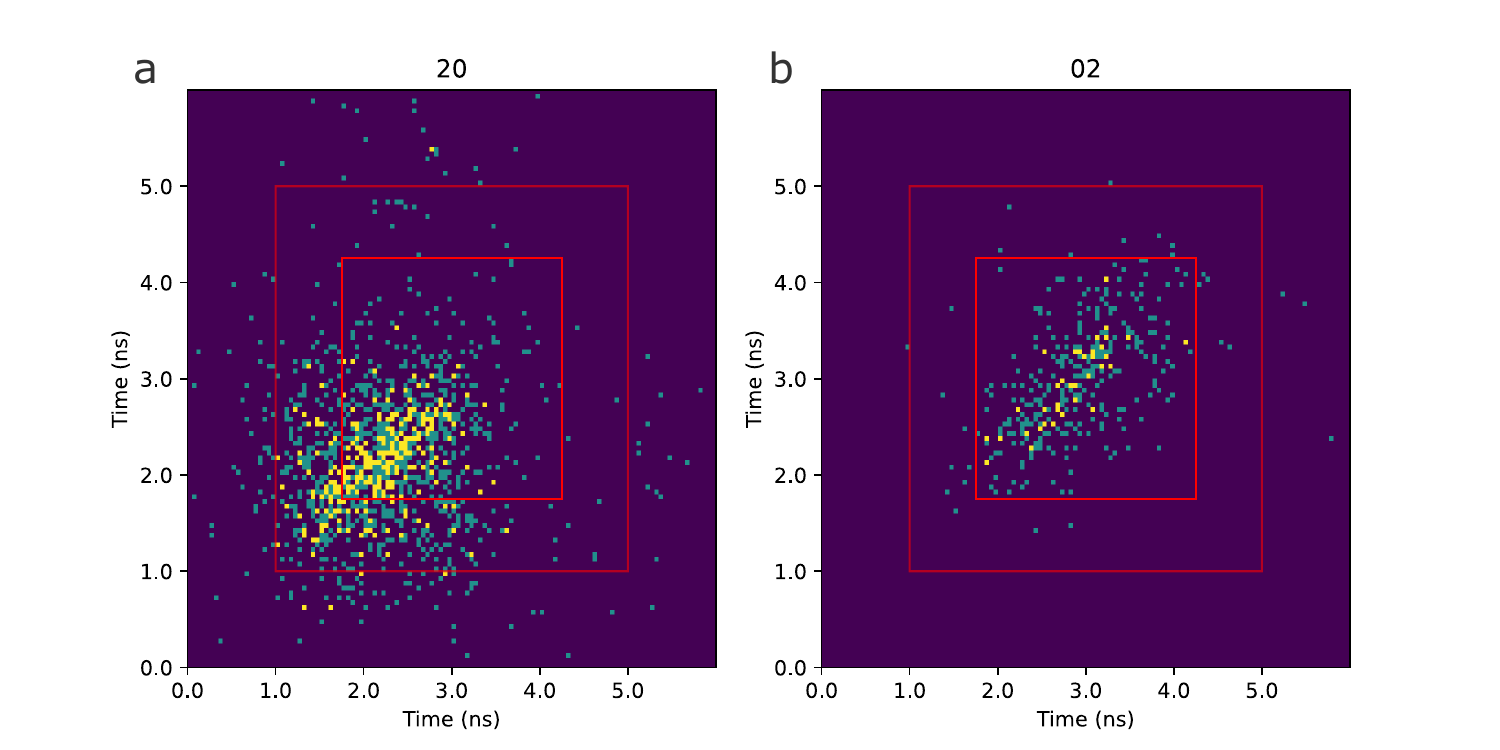}
    \caption{Joint temporal intensity distributions of two photons detected in the a. upper and b. lower mode of the photon sorting interferometer, labelled '20' and '02' in line with the probabilities in Fig.~\ref{fig:2}). The larger red box corresponds to a temporal window of $4\,\text{ns} \times 4\,\text{ns}  $ and was used for the unfiltered results in Fig.~\ref{fig:2}a,b. The smaller red box shows a temporal window of $2.5\,\text{ns} \times 2.5\,\text{ns}$ used for the filtered results in Fig.~\ref{fig:2}.}
    \label{fig:suppl:temporal_filtering}
\end{figure}

In Fig.~\ref{fig:2}d, we presented the results of photon sorting making use of a temporal filter. In this section, we provide more information. \newline
In Fig.~\ref{fig:suppl:temporal_filtering}, we show the joint temporal intensity (JTI) distributions detected at the different output modes of the interferometer.
For the filtered results in Fig.~\ref{fig:2}c, we only accept detections within the smaller red box corresponding to the temporal window of $2.5\,\text{ns}  \times 2.5\,\text{ns} $. As explained in the main text, the temporal filtering makes use of the fact that the elastic and inelastic parts of the scattered wavefunction are temporally different. The elastic part is mainly given by a two-dimensional Gaussian in the JTI, whereas the inelastic part is more elongated along the diagonal of the JTI. Using the temporal filter, we primarily discard photons from the elastic part of the scattering. 
This effectively corresponds to increasing the waveguide coupling $\beta$ and enhancing the nonlinearity strength.

\section{Single-sided, double-sided and chiral waveguides}
\label{sec:single_vs_chiral}

In the sections above, we use a chiral description of our system, which in practice is not a chiral waveguide. The waveguide used in this experiment is a single-sided waveguide terminated by a mirror on one side. This is however equivalent to a chiral waveguide where the emitter-waveguide coupling will be dependent on the distance to the mirror. See Fig. \ref{fig:chiral_vs_single} for a sketch of how the different situations look. In the following section, we will show this equivalence, which has also been derived for example in Ref.~\cite{fan2010input} and Ref.~\cite{witthaut_photon_2010}.
The Hamiltonian for the qubit is still the same. In the Hamiltonian for the photon fields, there is now a right and left propagating mode denoted with an \textit{r (l)} subscript for right (left) propagating modes, respectively.
\begin{equation}
    H_\text{ph}=\int\dd\omega a_{\omega,r}^\dagger a_{\omega,r} + \int\dd\omega a_{\omega,l}^\dagger a_{\omega,l} + \int\dd\omega b_\omega^\dagger b_\omega.
\end{equation}
The photon-emitter coupling Hamiltonian also needs revision. There will be a right and a left decay rate, which we assume are the same in the experiment. Upon reflection of the mirror, with mirror reflectivity $\eta$ and propagating the distance $x$ twice, the interaction is given by
\begin{equation}
    H_\text{qd-ph}=\sqrt{\frac{\gamma_r}{2\pi}}\int\dd\omega (\sigma_-a_{\omega,r}^\dagger + \text{h.c})+ \eta\sqrt{\frac{\gamma_l}{2\pi}}\int\dd\omega (\sigma_-a_{\omega,r}^\dagger e^{ik(\omega)2x} + \text{h.c})+\sqrt{\frac{\gamma_\text{loss}}{2\pi}} \int\dd\omega (\sigma_-b_\omega^\dagger + \text{h.c}).
\end{equation}
We will assume a constant dispersion $k(\omega)=k$, and that the emitter is much closer to the mirror than the lifetime of the emitter $2x \ll \frac{c}{n\gamma_l}$.
An ideal case would have a perfect mirror $\eta=1$, and the phase would be $kx=\pi$ to give constructive interference between the right and left propagating modes. If we assume $\gamma_l=\gamma_r$, this would result in an effective chiral system where the effective decay rate is twice that of the decay to either right or left, which is to be expected. For a phase of $kx=\pi/2$, destructive interference between the right and left propagating modes would occur, and the emitter would, in principle, never decay into the waveguide. Thus, by changing this phase, it is possible to modulate the effective decay rate into the waveguide and thus also the $\beta$-factor since this is the ratio between the total decay and the decay into loss channels. Losses due to an imperfect mirror can be added to the general loss channel, since it does not matter if photons are lost at the mirror. We can rewrite the Hamiltonian as 
\begin{equation}
    H_\text{qd-ph}=\int\dd\omega \left[\sigma_-a_{\omega,r}^\dagger \left(\sqrt{\frac{\gamma_r}{2\pi}}+\eta\sqrt{\frac{\gamma_l}{2\pi}}e^{i2kx}\right)+ \sigma_+a_{\omega,r} \left(\sqrt{\frac{\gamma_r}{2\pi}}+\eta\sqrt{\frac{\gamma_l}{2\pi}}e^{-i2kx}\right)\right]+\sqrt{\frac{\gamma_\text{loss}}{2\pi}} \int\dd\omega (\sigma_-b_\omega^\dagger + \text{h.c}).
\end{equation}
The new combined decay rates $\sqrt{\frac{\gamma_r}{2\pi}}+\eta\sqrt{\frac{\gamma_l}{2\pi}}e^{i2kx}$, will be complex, but the phase is a global phase which can be factored out. This means that the effective decay rate into the waveguide will be $\abs{\sqrt{\frac{\gamma_r}{2\pi}}+\eta\sqrt{\frac{\gamma_l}{2\pi}}e^{i2kx}}$. From this combined decay rate, the previously discussed constructive and destructive mirror-dependent interference is also clear. We can thus map a single-sided waveguide with a mirror to a perfect chiral waveguide, provided the mirror is ideal and yields constructive interference. In practice, imperfect mirror and emitter placement will reduce waveguide coupling, which can still be accounted for in the chiral model used in this experiment. 
In the modelling, we assume the mirror is perfect and the only loss channel is through t$\gamma_\text{loss}$. We also assume $\gamma_l=\gamma_r$.
%

\label{appendix:butterfly_comparison}
\begin{figure}
    \centering
    \includegraphics[width=\linewidth]{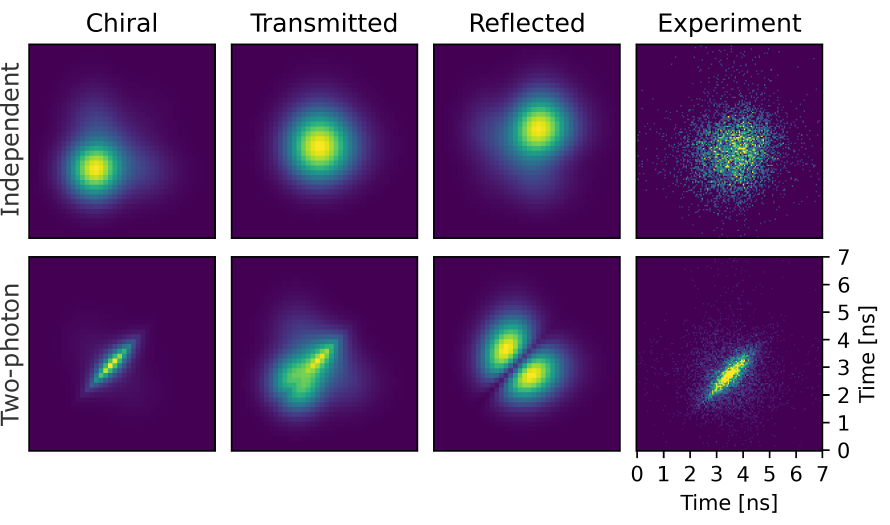}
    \caption{Joint temporal intensity distributions from one- and two-photon scattering events. Detection occurs directly after scattering from the QD. 
    Three different theoretical models are shown, corresponding to a perfect chiral (single-sided) waveguide, and the transmission and reflection ports of a double-sided waveguide, alongside experimental data for the single-sided waveguide. The first row corresponds to one-photon statistics, where we plot coincidence events between two individual (independent) photons from neighbouring pulses. The second row corresponds to the two-photon case, where we plot events from two (correlated) photons from the same pulse. All models use the same parameters for $\beta$ and $\sigma_\text{SD}$ as described in the main text. Note that the theoretical models assume no pure dephasing.}
    \label{fig:supp_CRT_distributions}
\end{figure}
The interaction between the emitter and optical field in a single-sided waveguide has been modelled as an effective chiral model as described in Appendix \ref{appendix:theory}. In this section, we further verify this claim by examining the temporal probability distributions of the scattered photons. Here, we directly detect scattered photons without interference in a time-bin interferometer. This experiment is similar to that of Ref.~\cite{jeannic2022dynamical}, but with a single-sided device. In Fig. \ref{fig:supp_CRT_distributions}, we compare the experimentally measured distributions to three different theoretical models. The theoretical models include a chiral model and a perfect non-chiral model, where we plot either the reflected or transmitted photons. The one-photon components are all very similar, with a pronounced Gaussian profile at the centre. However, the three theoretical two-photon distributions are very similar, and the experimental data clearly resemble those of the theory.

\section{Comparison to single emitter sorter architectures}
~\label{appendix:RTsorter}

There have been several demonstrations of photon sorting using an atomic system. Those approaches classify sorting capability by examining modifications to autocorrelation functions. 
For example, in Ref. \cite{bennett2016asemiconductor}, they perform photon sorting using a semiconductor quantum dot in a monolithic cavity. A weak pump creates a hole in the QD, thereby activating the optically addressable transition. Depending on the detuning of the incident field from the transition, they show bunching ($g^{(2)} >  1$) or anti-bunching ($g^{(2)} < 1$) in the reflected signal.

\begin{figure}
    \centering
    \includegraphics[width=\linewidth]{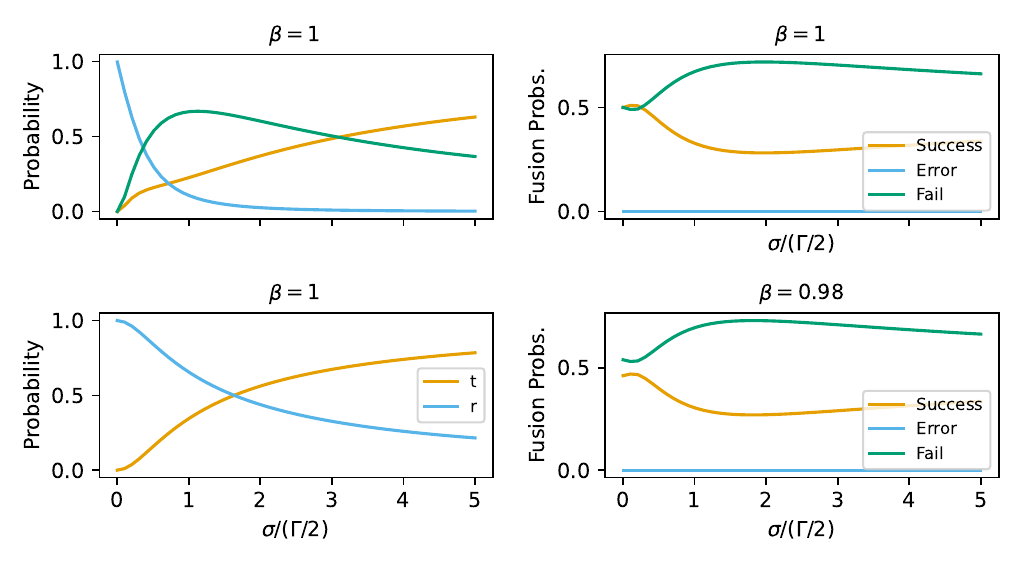}
    \caption{Sorting probabilities of a photon sorter solely based on a QD ideally coupled to a two-sided waveguide, as a function of the incoming photon pulse duration normalised to the quantum dot decay rate ($\sigma / \gamma$). 
    a. Probabilities for the two-photon state to be sorted into the transmission-transmission (tt, orange), reflection-reflection (rr, blue), and transmission-reflection (tr, green) modes. 
    b. Probabilities for a one-photon pulse to be transmitted (t, orange) or reflected (r, blue). Success, failure, and error probabilities for BSMs based on such a photon sorter, assuming (c) $\beta=1$ and (d) $\beta = 0.98$.}
    \label{fig:suppl:rt_photonsorter}
\end{figure}

In general, such an approach, involving a transition coupled to a two-sided waveguide, is not well suited to photon sorting, as the different number contributions are not effectively separated. While the modification of autocorrelation functions clearly indicates a change in the photon-number statistics due to the interaction, the actual probability with which the different photon-number contributions are separated into distinct paths remains unclear. \newline 
In the following, we will theoretically evaluate the performance of a photon sorter based on a two-sided QD. In Fig. \ref{fig:suppl:rt_photonsorter}a)-c), we show the results of numerical simulations (based on \cite{fan2010input,ramos2018correlated}) of an ideal two-sided QD perfectly coupled to a waveguide. We set $\beta=1$ (and correspondingly $\gamma_{loss} = 0$) as well as $\gamma_d = \sigma_{SD} = 0$ and plot the sorting probabilities as a function of $\sigma/(\Gamma/2)$, the incoming pulse width $\sigma$ normalised to the QD decay rate $\Gamma/2$. For $\sigma/(\Gamma/2) \ll 1$, most two-photon components are reflected but so are most one-photon components. At intermediate $\sigma/(\Gamma/2) \approx 1$, the majority of the two-photon components are split up into a reflected and a transmitted photon, while the one-photon component is partially transmitted, partially reflected. At higher $\sigma/(\Gamma/2)$, the two-photon component is either transmitted or split up, while the one-photon component is mostly transmitted. As is evident, there is no region in which the QD efficiently sorts the different contributions of numbers into separate spatial modes. \newline 
We can further illustrate this point by including this R/T sorter into the Bell State Measurement circuit from Fig. \ref{fig:1}c. We will identify the transmitted and reflected channels as the two separate rails in Fig. \ref{fig:1}e. These results are displayed in Fig. \ref{fig:suppl:rt_photonsorter}c, where we can identify a slim region where the success probability of the BSM exceeds 50$\%$ (the threshold set by linear optics), maxing out at 52$\%$. Thus, a BSM based on such an R/T sorter barely beats out regular linear optics. Moving towards a slightly non-optimal dot with $\beta = 0.98$ - around the highest $\beta$-factor ever demonstrated - as in Fig. \ref{fig:suppl:rt_photonsorter}d, the BSM already does not exceed a success probability of 50$\%$. 
Note that in these simulations, even for the dot with imperfect waveguide coupling in Fig. \ref{fig:suppl:rt_photonsorter}d, we assumed $\gamma_d = \sigma_{SD} = 0$. Realistic values for these parameters would further decrease the success probability of an R/T photon sorter-based BSM.

\section{Application: FBQC with photon-sorter-boosted fusions}
~\label{appendix:fusion}

For a general estimate of BSM performance, we assume that the linear optical circuits in the BSM operations (excluding the photon sorter optics) are ideal. The fusion is successful if both one-photon components of the $\ket{\psi^\pm}$ states are sorted correctly, given by probability $P_{10}^2$, or if the bunched $\ket{\phi^\pm}$ states are sorted correctly and not dephased, given by probability $P_{02}$ and $P_{02,\gamma_d}$ respectively. In the case of $\ket{\phi^\pm}$ being  dephased, we assume the photons are distinguishable from each other, and thus the detection output pattern will be random. These assumptions will set an upper bound on the fusion error, since, in practice, the photons will likely be partially distinguishable, leading to constructive interference in the correct modes. Assuming the photons are distinguishable, there are 4 modes in which each photon can be detected, yielding 16 equally likely outcomes. 4 of these result in a success, 4 result in error and the rest are detectable failures. Fusion errors can also occur if both photons of the $\ket{\psi^\pm}$ states are sorted incorrectly with a probability of $P_{01}^2$. Again, we assume the photons are distinguishable due to pure dephasing. This time, half the outputs will result in errors, and the other half in failures. The fusion failures will also happen from the remaining terms: $\ket{\phi^\pm}$ states incorrectly, giving rise to a bunching event at the first detection screen with probability $P_{20}^2$; $\ket{\phi^\pm}$ states that antibunch, giving rise to a single click at both screens with probability $P_{11}^2$; and $\ket{\psi^\pm}$ states where one is sorted correctly and the other incorrectly with probability $2P_{01}P_{10}$. From these considerations, it is possible to extract the expected fusion performance of the setup in Fig. \ref{fig:3} with this photon sorter to 

\begin{figure}
    \centering
    \includegraphics[scale=1]{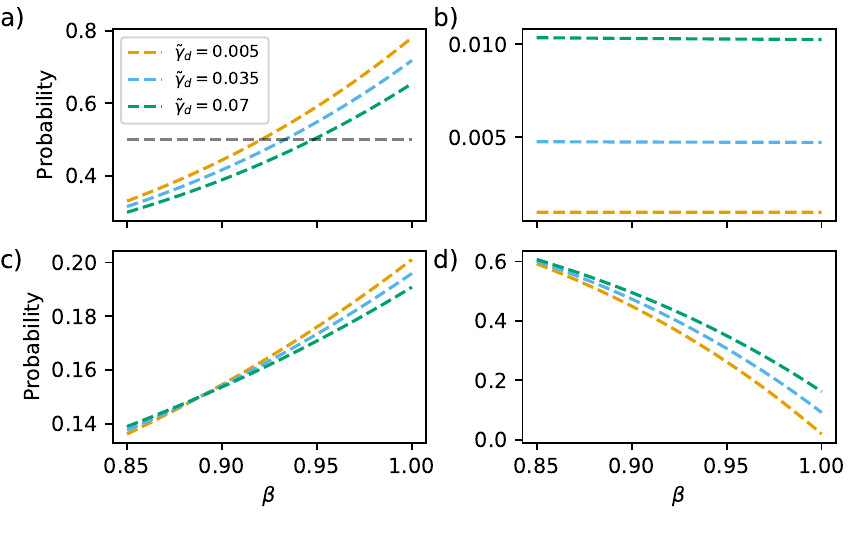}
    \caption{Fusion performance enhanced with a photon sorter. All rates are un-normalised, and thus a non-ideal $\beta$ factor result in losses. a) Success rate. b) Error rate. c) Failure rate. d) Loss rate.}
    \label{fig:fusion_probs_supp}
\end{figure}
Fusion-based quantum computing was introduced in Ref.~\cite{bartolucci2023fusion} and relies on two primitives:
\begin{enumerate}
    \item Small entangled resource states of constant size and structure. In our estimates, we use a linear chains from Ref. \cite{chan2025tailoring}, which are particularly suitable for quantum emitters.
    \item Some projective entangling measurement, called Fusion measurements, between individual qubits of two resource states. Most often, these are chosen as Bell state measurements (BSMs).
\end{enumerate}

\begin{figure}
    \centering
    \includegraphics[width=0.8\linewidth]{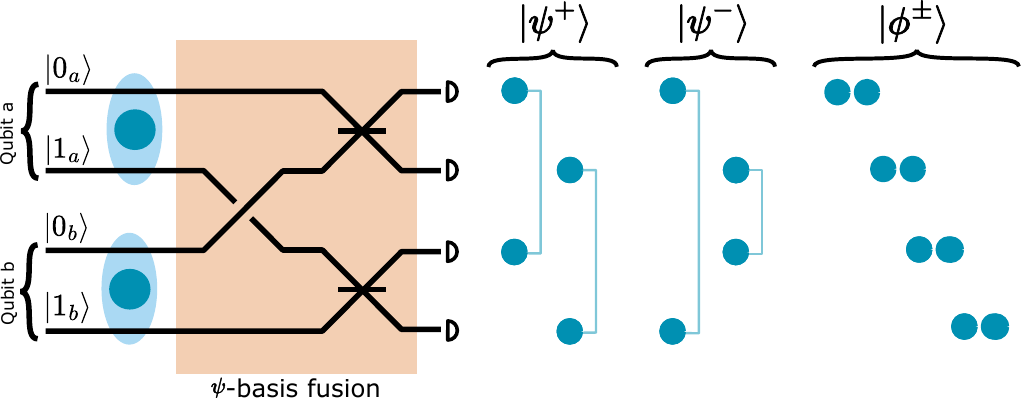}
    \caption{Linear-Optical BSM that implements a fusion in the $\phi$-basis. The $\ket{\phi^\pm}$ lead to unique distinguishable detector clicks, whereas the $\ket{\phi^\pm}$-states lead to bunching.}
    \label{fig:suppl:lo_bsm}
\end{figure}

The most common implementation of a BSM in the dual rail basis is via linear optics \cite{weinfurter_experimental_1994}, as illustrated in Fig. \ref{fig:suppl:lo_bsm}. First, $\ket{1}_a$ and $\ket{0}_b$ are swapped, and then the rails are sent pairwise into a 50:50 beam splitter. After the interference, a single-photon measurement is performed on all 4 rails. In this setup, we can calculate the statistics of the 4 Bell states and find that $\ket{\psi^{\pm}}$ yields unique detector-click combinations. The states $\ket{\phi^\pm}$ however will have their two photons bunch into the same detector, making the click patterns of $\ket{\phi^+}$ and $\ket{\phi^-}$ indistinguishable. If we had instead swapped $\ket{1}_a$ and $\ket{1}_b$, we could resolve the states $\ket{\phi^\pm}$, but the states $\ket{\psi^\pm}$ would lead to the same click patterns.

Thus the success probability is limited to $50 \%$ \cite{calsamiglia_maximum_2001, weinfurter_experimental_1994, braunstein_measurement_1995}. It can be increased to $75 \%$ by adding a pair of entangled photons \cite{grice2011arbitrarily} or even unentangled single-photon ancillae \cite{ewert2014efficient, hauser2025boosted}. \newline
An alternative approach is detailed in Fig. \ref{fig:1}a and uses a photon sorter - devices capable of sorting photons into different rails depending on the photon number.

To relate the photon sorting statistics from Fig. \ref{fig:2} to fusion success probabilities, we assume a lossless implementation of linear optics leading up to and following the photon sorters, as shown in Fig. \ref{fig:3}a. \newline
The success, failure and error probabilities averaged over all four Bell-states are then given by
\begin{align}
    P_{\text{success}} &= \frac{1}{2} \left( P_{10}^2 + P_{02} P_{02, \gamma_d} + \frac{1}{4} P_{02} (1 - P_{02, \gamma_d} )\right) \\
    P_{\text{failure}} &= \frac{1}{2} \left( 2 P_{01} P_{10} + P_V^2 + \frac{1}{2} \left( P_{01}^2 - P_V^2\right) + \frac{1}{2} P_{02} \left( 1 - P_{02, \gamma_d} \right) + P_{20} + P_{11} \right) \\
    P_{\text{error}} &= \frac{1}{2} \left(\frac{1}{2} \left( P_{01}^2 - P_V^2 \right) + \frac{1}{4} P_{02} \left( 1 - P_{02, \gamma_d} \right) \right) 
\end{align}
The factors of $\frac{1}{2}$ stem from the probability of having one of the states from $\ket{\psi^\pm}$ or $\ket{\phi^\pm}$, respectively. 
If the input states are $\ket{\psi^\pm}$, photons will antibunch before the photon sorters. Thus, the fusion is successful if both one-photon components entering different photon sorters are correctly sorted, as given by the probability $P_{10}^2$.
If the input states are $\ket{\phi^\pm}$, the photons will bunch before the photon sorters, and the fusion will be successful if the photons are sorted correctly and not dephased, which happens with probability $P_{02}  P_{02, \gamma_d}$. If the photons are properly sorted but dephased, we can assume they are distinguishable. In this case, they will be distributed individually across the detectors with equal probability. Out of the 16 possible combinations of two photons to occupy four detectors, 4 of these will lead to the correct click pattern, contributing the term $\frac{1}{4} P_{02} (1 - P_{02, \gamma_d} ) $ to the success probability. Four other combinations yield an incorrect click pattern (misassigning $\ket{\phi^+}$ to $\ket{\phi^-}$ or vice versa), contributing the same term to the error probability. The remaining eight out of 16 cases generate an unassigned click pattern and are classified as failures.

Apart from the case above, a Bell state can be misclassified as $\ket{\psi^\pm}$ if both photons are individually sorted incorrectly.
In this case, if the photons are sorted incorrectly due to a static imbalance in the photon sorter interferometer (with probability $P_V^2$), they remain indistinguishable and will bunch on the second detector screen, leading to a failure. Otherwise, we assume the photons are distinguishable. They will distribute uniformly over the second detector screen, leading to an assigned detector click pattern half of the time ($\frac{1}{2} ( P_{01}^2 - P_V^2)$). 

Fusion failures occur in a few cases. If the state is $\ket{\psi^{\pm}}$, we identify a failure if one of the two one-photon components is sorted incorrectly (with probability $2 P_{01} P_{10}$), if both photons are sorted incorrectly due to a static imbalance of the interferometer and the resulting click pattern is not assigned to a Bell state ($\frac{6}{10} P_V^2$), if both photons are sorted incorrectly due to pure dephasing and result in an unassigned click pattern ($\frac{1}{2} ( P_{01}^2 - P_V^2)$). If the state is $\ket{\phi^\pm}$, we identify a failure if the two-photon state is sorted correctly but dephased in the interaction and leads to an unassigned detector pattern ($\frac{1}{2} P_{02} (1 - P_{02, \gamma_d})$) or simply if the two-photon state is sorted incorrectly; either both photons into the top mode ($P_{20})$ or one photon into each mode ($P_{11}$).

\section{Application: Quantum key distribution using photon sorter}
\label{appendix:qkd}

One possible application of the photon sorter BSA is entanglement swapping, e.g., for long-range entanglement distribution.
For approaches like the DLCZ protocol \cite{duan_long-distance_2001} it is fundamentally required, but also in matter platforms it becomes necessary if many qubits are multiplexed such that all-to-all connectivity becomes feasible only when mediated by photons.

To generate entanglement over a long distance $L$, one usually splits the distance into $2^n$ segments by using repeaters. In this case, we express the entanglement generation rate for the outer links as \cite{sangouard_quantum_2011}
\begin{equation}
    R = R_f \left( \frac{2}{3} P_S \right)^n
\end{equation}
where $R_f$ is the rate to generate entanglement in a fundamental link,
which can usually be modelled as $R_f = R_0 ({c 2^n}/{L}) \eta (1-F)$ with the trial time ${c 2^n}/{L}$, speed of light $c$, transmission efficency $\eta$, fidelity $F$, as well as a protocol dependent factor $R_0$. 
The transmission efficiency for central heralding stations $\eta = \eta_I e^{- L / (2^{n+1} L_{\text{att}})}$ can be split into an intrinsic efficiency $\eta_I$ and fiber losses with an attenuation length $L_{\text{att}}$.
The second part of the rate accounts for the swap probability $P_S$, and a $2/3$ factor for parallelisation \cite{sangouard_quantum_2011}.
From this, it is immediately evident that increasing $P_S$ will accelerate entanglement generation:
 using only linear optics, an optical entanglement swap the  success probability is limited to  $P_S=0.5$. For the implementation proposed in \cite{witthaut2012photon}, $P_S=0.86$ can be achieved with a perfect quantum emitter, which rises to $P_S=1$ for a perfect photon sorter.

For a fair and concrete comparison, we will include errors introduced in the swap and focus on the secret key rate of an entanglement-based quantum key distribution protocol.
We consider an asymmetric six-state protocol where we can express the secret key rate as $K = R p f$ in terms of the entanglement generation rate $R$, the sifting probability $p \approx 1$, and the secret fraction $f = 1 - h(Q) - Q - (1-Q) h \left( \frac{1 - 3 Q/2}{1 - Q} \right)$, using the binary entropy $h(Q)$ \cite{scarani_security_2009}.
For a Werner state
\begin{equation}
\rho = (1 - \epsilon) |{\phi_+}\rangle \langle{\phi_+}| + \frac{\epsilon}{4}  \mathbb{I} ,
\end{equation}
with fidelity $F = 1 - {3 \epsilon}/{4}$, the quantum bit error rate is $Q = {2} (1-F) / {3} = {\epsilon}/{2}$ \cite{scarani_security_2009}.
Considering that any two-qubit state can be transformed to a Werner state using twirling operations \cite{bennett_mixed-state_1996}
we will restrict the following analysis to Werner's states and note that the achieved rate represents a lower limit.

Assuming that the initial entanglement generation happens with an infidelity $1-F_0 = {3 \epsilon_0}/{4}$.
We are interested in the total error of a Bell pair after performing $n$ layers of swaps.
To this end, we use the error probability ${3 \epsilon}/{4}$ of the Bell state analyser, i.e., the probability that the BSA randomly falsely heralds a false Bell state.
This can be modelled as a decoherence channel after the ideal BSA.
Up to linear order in the errors, we can add the error of the previous level and the entanglement swap errors when increasing the swap level,
such that the error after $n$ swap levels is $1-F_n = {3 \epsilon_n}/{4}$ with $\epsilon_n = 2^n \epsilon_0 + (2^n - 1) \epsilon$.
Note that this is the solution of the recurrence relation $\epsilon_{n+1} = 2 \epsilon_n + \epsilon$.
With this error, we can estimate the secret fraction $f$ and, together with $R$, the secret key rate $K$.

For concreteness, we also determine an example for $R_0$ in which each node has $N$ qubits ($2N$ for repeater stations).
Each qubit is prepared in a qubit-photon state $\sqrt{1-p} \ket{0}_q \ket{0}_p + \sqrt{p} \ket{1}_q \ket{1}_p$ with emission probability $p$, qubit (photon number) state $\ket{k}_{q(p)}$ for $k=0,1$.
The photons are sent to the central heralding station, where they are pairwise combined on a balanced beam splitter, and entanglement between the qubits is heralded upon the detection of a single click at the output ports of the beam splitter.
Using non-photon-number resolving detectors at the heralding station, a single attempt has a success probability of $2 p \eta$ and a Fidelity of $F_0 = 1 - p$.
For low success probabilities, the probability that any of the $N$ multiplexed links succeeds is $2 N \eta (1-F_0)$, such that we find $R_0 = 2 N$.

We display the secret key rate as a function of distance (for the concrete example of $R_0 = 2N = 200$) 
in the main text.

Therein, we compare the perfect non photon sorted entanglement swap with $P_S =0.5, \epsilon = 0$, a state-of-the-art quantum emitter with $P_S \sim 0.645, \ \epsilon \sim 0.004$ and the best possible implementation of our method using perfect quantum dots $P_S = 0.86, \epsilon = 0$ \cite{witthaut2012photon}. We note that the most limiting factor towards a high (non post-selected) photon sorting success probability is the waveguide coupling $\beta$, which reduces the sorting fidelity due to photons scattered out of the waveguide. For example, the quantum emitter used in the lab has post-selected probabilities of $P_S = 0.575, \epsilon \sim 0.012$ and the state-of-the-art QD has $P_S \sim 0.69, \epsilon \sim 0.004$ \cite{arcari2014near}.
Above, we used the error introduced by the swap $\epsilon = 4 P_{\mathrm{err}} / (3 P_S)$ and the total success probability $P_S = P_{\text{succ}} + P_{\text{err}}$ as a function of the the probability to succeed $P_{\text{succ}}$ and the error probability $P_{\text{err}}$.

\end{document}